\documentstyle[12pt]{article}
\begin{document}

\setlength{\baselineskip}{15.2pt}

\newcommand\ie {{\it i.e. }}
\newcommand\eg {{\it e.g. }}
\newcommand\etc{{\it etc. }}
\newcommand\cf {{\it cf.  }}
\newcommand\etal {{\it et al. }}
\newcommand{\be}{\begin{eqnarray}}
\newcommand{\ee}{\end{eqnarray}}
\newcommand\Jpsi{{J/\psi}}
\newcommand\M{M_{Q \overline Q}}
\newcommand\mpmm{{\mu^+ \mu^-}}
\newcommand{\jp}{$ J/ \psi $}
\newcommand{\pp}{$ \psi^{ \prime} $}
\newcommand{\ppp}{$ \psi^{ \prime \prime } $}
\newcommand{\dd}[2]{$ #1 \overline #2 $}
\newcommand\noi {\noindent}

\begin{flushright}
LBL-37981
\end{flushright}
\vspace{1cm}

\begin{center}
{\Large How To Find Charm in Nuclear Collisions at RHIC and LHC}\\[5ex]
\alph{footnote}

S. Gavin,$^{a,}$\footnotemark P. L. McGaughey,$^b$ P. V. Ruuskanen,$^c$ and
R. Vogt$^{d, e,}$\footnotemark \\[4ex]
\addtocounter{footnote}{-1}
\footnotetext{This manuscript has been authored under contract number 
DE-AC02-76CH00016 with the U. S. Department of Energy.  Accordingly, the 
U.S. Government retains a non-exclusive, royalty-free license to publish
or reproduce the published form of this contribution, or allow others to 
do so, for U.S. Government purposes.}
\addtocounter{footnote}{1}
\footnotetext{This work was supported in part by the Director, 
Office of Energy Research, Division of Nuclear Physics
of the Office of High Energy and Nuclear Physics of the U. S.
Department of Energy under Contract Number DE-AC03-76SF0098.}

$^a$Physics Department, Brookhaven National Laboratory, Upton,
NY 11973\\

$^b$Los Alamos National Laboratory, Los Alamos, NM 87545\\

$^c$Department of Physics, University of Jyv\"{a}skyl\"{a},
Jyv\"{a}skyl\"{a}, Finland\\

$^d$Nuclear Science Division, Lawrence Berkeley National Laboratory,
Berkeley, CA 94720\\

$^e$Physics Department, University of California at Davis, Davis, 
CA 95616\\

\end{center}

\vspace{1cm}
\begin{center}
{\bf Abstract}
\end{center}
\begin{quote}
\begin{small}
Measurements of dilepton production from charm decay and Drell-Yan
processes respectively probe the gluon and sea quark distributions in
hadronic collisions.   In nucleus-nucleus collisions, these hard scattering
processes constitute a `background' to thermal contributions from  the
hot matter produced by the collision.   To determine the magnitude and
behavior  of this 
background,  we calculate the hard scattering contribution to dilepton 
production in nuclear collisions at RHIC and LHC at  next to leading order in 
perturbative QCD.  Invariant mass, rapidity and transverse momentum 
distributions 
are presented.  We compare these results  to optimistic hydrodynamic 
estimates of the thermal dilepton production.  We find that   charm production
from hard scattering  is by far  the dominant contribution.  
Experiments  therefore can measure   the gluon 
distribution in the nuclear target and projectile and, consequently,
can provide new information on gluon shadowing.  
We then illustrate how  experimental cuts on the rapidity gap between the 
leptons can aid  in reducing the charm background, thereby enhancing thermal
information.\\[2ex]
\end{small}
\end{quote}

\newpage
\setcounter{footnote}{0}
\section{Introduction}

Dilepton production provides an important tool for measuring the temperature  
of the high density matter produced in the early stages of a relativistic 
heavy-ion collision (see \cite{VJMR} and references therein).  To make use 
of this tool at ion colliders such as RHIC and LHC,  we must understand  
additional sources of dileptons from hard scattering and other 
non-equilibrium processes.  In high energy  
$pp$ interactions,  the continuum in the dilepton mass range above $M\sim 
2$~GeV is dominated by the Drell-Yan process and by semileptonic decays of 
$D$ and other charm mesons.  These leptons are produced by hard scatterings, 
at scales exceeding $M$ and $2m_{c}\sim 3$~GeV 
respectively, so that their production can be addressed using perturbative 
QCD. In contrast, the lower mass region arises from soft 
processes, for which theory offers little guidance.  We therefore hope 
to find a signal of thermal dilepton production at masses $M\sim 2-3$~GeV 
where {\it i}) the thermal contribution can still be sizable and {\it ii}) 
the background is calculable.  

In this paper, we compare the predictions of dilepton production from
a simple thermal model with the hard `background' from heavy quark
pair, $Q \overline Q$, decays and Drell-Yan production at RHIC and LHC
nucleus-nucleus collision energies, $\sqrt{s} = 200$ GeV and 5.5 TeV
in the nucleon-nucleon center of mass.  Previously we compared the
thermal dilepton and thermal charm rapidity distributions with
Drell-Yan and initial charm production \cite{VJMR} calculated at
leading order (LO).  Our new next-to-leading order (NLO) results,
shown to agree with $pp$ and $p\overline p$ data in \cite{hpdy,hpoc},
no longer depend on arbitrary phenomenological `K factors' (often
incorrectly taken to be `2' in the literature).  The remaining
uncertainties in the NLO perturbative approach are well defined and
likely rather small \cite{hpdy}.  Also new to this paper are
calculations of the invariant mass and transverse momentum
distributions of the lepton pairs.  We use these distributions to
illustrate how thermal and hard dileptons can be distinguished in an
experiment.  In addition, we introduce the dilepton contribution from
$B$ meson decays, and discuss the effects of nuclear shadowing on the
initial production.

Figure 1 shows the invariant mass distributions of the calculated
contributions to the dilepton continuum in central nucleus-nucleus
collisions at RHIC and LHC.  We find that hard charm quark production
and decay dominates the continuum below the $\Upsilon$ mass.  In
particular, the charm signal is more than an order of
magnitude above the optimistic thermal dilepton and thermal charm
rates for $M>2$ GeV.  This result implies that dilepton measurements
can be used to extract the low $x$ gluon density in the nucleus (see
also \cite{LG2}).  On the other hand, the isolation of thermal signals
will not be straightforward.

Additional complications can arise from the fact that the charm
production cross section is large enough for multiple $D \overline D$
pairs to be produced in a single nucleus-nucleus collision.
Uncorrelated pairs form when a lepton $l^{+}$ from one $D\overline D$
is randomly paired with a $l^{-}$ from another $D\overline D$; the
correlated signal in Fig.\ 1 includes only dileptons from $D\overline
D$ pairs in which both quarks decay to leptons.  In central Au+Au
collisions at RHIC, up to 67 uncorrelated pairs may be produced while
at the LHC, over 3000 uncorrelated pairs will contribute to the raw
continuum from Pb+Pb collisions.  Uncorrelated charm does not affect
the measurement of parton densities, but it further complicates the
task of extracting thermal information.  Ideally the uncorrelated
pairs can be removed by a like-sign subtraction, leaving only the
pairs shown in Fig.\ 1.  Whether this subtraction can hold at the
accuracy needed to extract thermal signals is another matter.

To enhance the thermal signal in an experiment, one can choose to count
only those lepton pairs that have a small separation in rapidity.  As
observed by Fischer and Geist, dilepton pairs from charm decays 
typically occur with a large rapidity gap \cite{FiGe}.  No such gap is
present in Drell-Yan pairs or -- importantly -- in thermal pairs. 
Together with like sign subtraction, we expect  that rapidity gap cuts 
can essentially  remove 
the uncorrelated charm contamination and greatly suppress the correlated charm 
background.  Indeed, the finite acceptance of a real detector can serve 
a similar purpose.  We show that the acceptance window of PHENIX/RHIC and 
ALICE/LHC  can  enhance the signal from thermal charm decays 
to the point of measurability by rejecting  pairs with large  gaps. 

This paper is organized as follows: In section 2, we discuss in detail
Drell-Yan, $D \overline D$, and $B \overline B$ production in $pp$
interactions and comment on how these results may be modified by
nuclear shadowing.  In section 3, we show the results for thermal
dilepton and thermal $D \overline D$ pair production at RHIC and LHC,
assuming the most optimistic scenario for the initial conditions to
maximize the thermal rate.  We compare the initial hard production of
dileptons to the thermal model results in section 4.  We discuss the
effects of realistic detector geometries on the dilepton spectra at
RHIC and LHC and draw our conclusions.

\section{ Dilepton Production from Initial Interactions}

Perturbative QCD calculations of Drell-Yan and heavy quark production
at leading order have long been available.  The LO calculations
differed from the experimental measurements by a $K$-factor ($K =
\sigma^{\rm exp}/\sigma^{\rm theory}$) of 2-3 for charm production and
1.5-2 for Drell-Yan production.  This difference suggested that higher
order corrections to the production cross sections were important.
Additionally, while multiplying the leading order cross section by a
$K$ factor describes the single-inclusive quark distributions as well
as the mass and rapidity distributions of Drell-Yan and $Q \overline
Q$ pairs, the pair $p_T$ distributions for both processes are trivial
at leading order since the pairs are produced back-to-back.  Therefore
next-to-leading order calculations are necessary to fully describe
hard dilepton production.

In our previous work \cite{VJMR} we used the LO cross sections.  
Since then, an NLO   treatment of $Q \overline Q$ production has been 
made available \cite{MNR}.   An NLO treatment of the  Drell-Yan
$p_T$ distribution is also now available  \cite{hpdy,AK}.
With the NLO description of the perturbative cross sections,
we also use NLO evaluations of the parton densities.  Our results are
obtained using the MRS D$-^\prime$ \cite{D0}
parton densities\footnote{All available parton
distribution functions are contained in the package PDFLIB \cite{PDF},
available in the CERN library routines.}, compatible with the low $x$ data
from HERA \cite{HERA}.  The MRS D$-^\prime$ sea quark and gluon distributions
grow $\propto x^{-1/2}$ at the initial
scale, $Q_0^2 = 5$ GeV$^2$, when $x \rightarrow 0$.  
Using recent parton distribution functions that
agree with the HERA data produces a substantial increase over our previous
results \cite{VJMR}, obtained with obsolete leading order parton distributions
that become constant at $Q_0^2$ as $x \rightarrow 0$.  
Observe  that  older parton
distributions such as \cite{DO}  substantially underestimate the 
initial Drell-Yan and $Q \overline Q$ production at heavy-ion colliders. 

We remark that updated versions of the MRS distributions are available, 
including 
MRS G \cite{MRSG} which has a slower low $x$ growth than MRS D$-^\prime$.
While small, the changes from D$-^{\prime}$ to G  affect our charm results
most strongly. Compared to D$-^{\prime}$, we find that charm rates 
from G are  5$\%$ larger at RHIC and 20$\%$ smaller at LHC.  No doubt,
these numbers will continue to improve at that  level as more data
is analyzed, both from HERA and from Fermilab \cite{E665}.

While the NLO evolution generally improves the agreement between the theory and
the data, additional uncertainties are introduced,
including dependence on scale and scheme.
At fixed order in perturbation theory, the calculations depend on the
renormalization scale and the factorization scale as
well as the regularization scheme ($\overline{\rm MS}$ or DIS).
The hard scattering matrix elements and the
definition of the parton densities for each process are specified by the
regularization scheme.  In this paper, we have used the $\overline{\rm MS}$
scheme.  The renormalization scale enters into the strong coupling constant,
$\alpha_s$, and the partonic cross sections while the parton distributions
are evaluated at the factorization scale.  The precise relationship between
these scales and the momentum transfer, $Q$, is not uniquely defined.
However, since the parton densities are analyzed assuming that
they are equivalent, we also assume this.
If the perturbative expansion converges, further higher-order corrections
are small at large enough values of the scale $\mu$.  
For such scales the physical
cross section should become independent of the scale and scheme when
calculated at higher and higher orders. If the $\mu$ dependence is strong,
the perturbative calculation at that order is unreliable and further
higher-order corrections are necessary \cite{Ellis}.
The rates from the initial hard scatterings are rather sensitive to
the scale and scheme in the moderate mass and $p_T$ regime relevant to
heavy-ion experiments.

Another uncertainty in our results
involves the nuclear dependence of hard processes.
When the charged parton distributions are probed in deep-inelastic scattering
with a nuclear target and compared to a deuterium target, the ratio $R_{F_2} =
F_2^A/F_2^D$
has a characteristic shape as a function of $x$.  A depletion in the nucleus is
observed at low $x$, the shadowing region, and intermediate $x$, the EMC
region.  Shadowing occurs in the region below $x\sim 0.1$ while
the EMC region refers to $0.3<x< 0.7$.  Between the shadowing and EMC regions,
$R_{F_2}>1$, referred to as antishadowing.  Although the origin of this
behavior is not well understood, it is postulated to be
either an interplay of coherent and incoherent multiple scatterings in the
target or a modification of the parton densities in nuclear matter.  In any
case, the effect can be modeled phenomenologically
by a parameterization to fit the nuclear
deep-inelastic scattering data and implemented by a modification of the parton
distributions in the nucleus.  In the appendix, we choose two different
parameterizations of the nuclear parton densities to illustrate the effect in
$AA$ collisions at
RHIC and LHC.  The first is a general fit to the most recent nuclear target
data \cite{EQC} that does not differentiate between quark, antiquark, and gluon
modifications and does not include evolution in $Q^2$.  The second modifies the
valence and sea quark and gluon distributions separately and includes $Q^2$
evolution \cite{KJE} but is based on a fit to somewhat older 
data using obsolete
parton densities.  We find that at RHIC energies the charm and Drell-Yan
yields may be reduced 40-50\% while the bottom yield is changed by 
$\approx 10$\%.  
At the LHC, all the yields are reduced 50-60\%.  The results depend
on the $x$ region probed and, in the case of the second parameterization, the
scale $Q^2$.

\subsection{ Drell-Yan Production}

A detailed discussion of Drell-Yan production in high-energy $NN$ collisions
can be found in Ref.\ \cite{hpdy}.  We repeat some of the pertinent points
here. At leading order,
Drell-Yan pairs are produced by $q \overline q$ annihilation
into a virtual photon which decays to a lepton pair, $q \overline q \rightarrow
\gamma^\star \rightarrow l^+ l^-$.  The production cross section for
lepton pairs with invariant mass, $M$, and rapidity $y$, summed over quark
flavor $f$
is \be M^2 \frac{d\sigma_{\rm DY}}{dy dM^2} = 
\widehat{\sigma}_0 \tau \sum_f e_f^2 [
q_f(x_1,\mu) \overline q_f (x_2,\mu) + (1 \rightarrow 2) ] \, \, , \ee
where $\widehat{\sigma}_0 = 4\pi \alpha^2/9M^2$ is the LO $q
\overline q$ annihilation cross section, $\tau = M^2/s$, $\mu$ is the scale,
and $x_{1,2} = \sqrt{\tau} e^{\pm y}$ are the projectile and target momentum
fractions at which the parton densities are evaluated.
For $y>0$, $x_1$ increases with rapidity while $x_2$ decreases.
At NLO, the Compton and
annihilation processes $q g \rightarrow q \gamma^\star$ and $q \overline q
\rightarrow g \gamma^\star$ contribute in addition to vertex corrections to the
LO cross section so that eq.\ (1) is replaced with
\be  M^2 \frac{d\sigma}{dy dM^2} & = & \widehat{\sigma}_0
\tau \int_0^1 dx_1 dx_2
dz \delta(x_1 x_2 z - \tau) \delta \left( y -
\frac{1}{2} \ln\left( \frac{x_1}{x_2}
\right) \right) \\ \nonumber & & \hbox{} \times \left\{ \left[ \sum_f  e_f^2 [
q_f(x_1,\mu) \overline q_f (x_2,\mu) + (1 \rightarrow 2)] \right] \left[
\delta(1-z) + \frac{\alpha_s(\mu)}{2\pi} f_q(z) \right] \right. \\ \nonumber
&    & \hbox{} + \left. \left[ \sum_f e_f^2 [g(x_1,\mu)(
q_f(x_2,\mu) + \overline q_f (x_2,\mu)) + (1 \rightarrow 2)] \right] \left[
\frac{\alpha_s(\mu)}{2\pi} f_g(z) \right] \right\} \, \, . \ee  Note that
going to NLO requires a redefinition of $x_1$ and $x_2$
since the unobserved parton contributes to the total momentum in the final
state.  The correction terms $f_q$ and $f_g$ are
regularization scheme dependent.  The scale and scheme dependences are not
large for $M > 4$ GeV.  We will only consider Drell-Yan production for $M\geq
2$ GeV since below this value the perturbative calculation becomes unreliable.
The mass and rapidity distributions were calculated using a program provided 
by Rijken and van Neerven \cite{vNR}.

The $p_T$ dependence is trivial
at LO --- the lepton pair has $p_T=0$ if no intrinsic parton $p_T$ is
included.
Some of the Drell-Yan pair transverse momentum can be accounted for by
introducing a soft intrinsic $p_T$ distribution with 
$\langle p_T \rangle \sim 0.3$
GeV.  However, this is too small to account for the measured
$p_T$ distributions with $\langle p_T \rangle \sim 1$ GeV at $M=10$ GeV.  At
NLO, the pair acquires transverse momentum through recoil of the virtual
photon with the final-state parton.  Here the
$p_T$ dependence can be calculated perturbatively and is well behaved for
$p_T \sim M$.  However, at low $p_T$, the perturbative expansion parameter,
$\alpha_s
\ln^2 (M^2/p_T^2)$, becomes large and the expansion breaks down, making it
necessary to resum the perturbation series.  The resulting
cross section is \be M^2 \frac{d\sigma_{\rm resum}}{dp_T^2 dy dM^2} = \pi
\widehat{\sigma}_0 \tau e_q^2 \int \frac{d^2b}{(2\pi)^2} e^{i {\bf b \cdot
p_T}} \, W(b) \, \, , \ee where a Fourier transformation is made to impact
parameter space and the form factor, $W(b)$, effectively sums the leading and
sub-leading logarithms.
The $p_T$ distributions have been
calculated using a code developed by Arnold and Kauffman for $p \overline 
p\rightarrow W, Z$  \cite{AK} and extended to Drell Yan in $pp$ collisions 
in \cite{hpdy}.  This code
includes a method of interpolating between the low and high $p_T$ behavior.

The number of Drell-Yan pairs of mass $M$
produced in central nucleus-nucleus collisions, \be
\frac{dN_{\rm DY}}{dM} = T_{AB} (0) \frac{d\sigma_{\rm DY}}{dM} \, \, , \ee
is always small in the
range where the perturbative calculation is expected to be applicable,
{\it i.e.} $M>2$ GeV.  However, only a calculation of the mass-integrated cross
section can reveal whether or not the $AB$ rate is large enough for false
Drell-Yan pair production to be a problem.  
Since such a calculation is not possible over all masses, 
experimental techniques must be 
used to insure that these uncorrelated Drell-Yan
pairs do not contaminate the spectrum.  For example, a $p_T$ cutoff should 
reduce the probability of false high mass
Drell-Yan pairs arising from low mass pairs 
with large rapidity separation.  Additionally, a study of the angular 
distribution would also reduce any uncorrelated Drell-Yan pairs.
The number of lepton pairs in central Au+Au
collisions at $M=2$, 4, and 6 GeV are given in Table 1 for RHIC and in Table 2
for central Pb+Pb collisions at the LHC.  They are
significantly larger than in our previous work \cite{VJMR} due to the
different small $x$ behavior of the parton densities.  The average mass and
$p_T$ of the Drell-Yan pairs (with $M>2$ GeV) produced at RHIC and LHC 
energies are shown in Table 3.

Figures 2 and 3
show the NLO rates for Au+Au collisions at RHIC and Pb+Pb collisions at the
LHC. The calculated $p_T$ distributions for $M=4$ and 6 GeV at $y=0$ and 2 are
shown in Figs.\ 2(a) and 3(a).  The distributions at higher rapidity are
nearly parallel until the kinematic limit is approached.  Given this, we 
have assumed
that the $p_T$ distribution has the same shape for $M=2$ GeV and scaled
the 4 GeV results to obtain the $p_T$ distribution at $M=2$ GeV.
The rapidity distributions
are shown in Fig.\ 2(c) for RHIC energies and 3(c) for the LHC.  The cross
section grows with rapidity until the kinematic limit is approached because
$x_2 \overline q (x_2, \mu)$ increases and $x_1 q (x_1, \mu)$ decreases with
increasing $y$.
As previously discussed, the parton density dependence is enhanced for low
masses and high rapidities at $\sqrt{s} = 5.5$ TeV even though the
rapidity-integrated results do not depend strongly on the parton
densities.  Note that the strongest dependence on the parton densities occurs
outside the range of current measurements.  In our previous work,
the variation with rapidity was weaker because we used obsolete
parton distributions where $x \overline q(x 
\rightarrow 0,Q_0^2) \rightarrow {\rm constant}$.

\subsection{Heavy Quark Production and Decay}

At RHIC and LHC energies, heavy quark production will be substantial.
We consider both charm and bottom production and fragmentation into heavy
mesons which subsequently decay to lepton pairs.
The double differential heavy meson pair
production cross section at LO is \cite{VBH2,Ban}
\be E_H E_{\overline H} \frac{d \sigma_{H \overline H}}{d^3p_H
d^3p_{\overline H}} & = &
\int \frac{\hat{s}}{2 \pi} \frac{dx_1}{x_1} \frac{dx_2}{x_2} dz_H
dz_{\overline H}
C(x_1,x_2) \frac{E_H E_{\overline H}}{E_Q E_{\overline Q}} \\ \nonumber &   &
\frac{D_{Q/H}(z_H)D_{\overline Q/\overline H}(z_{\overline 
H})}{z_H^3 z_{\overline H}^3}
\delta^4 (p_1 + p_2 - p_Q - p_{\overline Q}) \, \, , \ee
where $H(Q \overline q)
\overline H(\overline Q q)$ is the heavy meson pair, $D \overline D$ for
$Q = c$ and $B \overline B$ for $Q=b$.
The fragmentation function, $D_{Q/H}(z)$, 
describes the hadronization of the heavy quarks where $z=p_H/p_Q$ is the
fractional momentum of the heavy quark carried by the hadron.  
Charm hadroproduction at low $p_T$ is best described by the assumption that
the charmed quark experiences no momentum loss during hadronization,
{\it i.e.}\ $D_{Q/H}(z) = \delta(1-z)$ \cite{VBH2}, resulting in
a shift between $y_Q$ and $y_H$ due to the quark and meson mass difference.
At LO heavy quarks are produced by gluon fusion, $g g \rightarrow Q
\overline Q$, and quark-antiquark annihilation, $q \overline q \rightarrow Q
\overline Q$.  To LO, the convolution function $C(x_1,x_2)$ is \be
C(x_1,x_2) & = & \sum_f [x_1 q_f(x_1,\mu) x_2\overline q_f(x_2,\mu) +
x_1 \overline q_f(x_1,\mu) x_2q_f(x_2,\mu)]
\frac{d \widehat{\sigma}(x_1,x_2,\mu)}{d \widehat{t}}|_{q \overline q} \\
\nonumber &   & \hbox{} + x_1g(x_1,\mu) x_2g(x_2,\mu)
\frac{d \widehat{\sigma}(x_1,x_2,\mu)}{d \widehat{t}}|_{gg} \, \, , \ee
where $d \widehat{\sigma}/d
\widehat{t}$ can be found in {\it e.g.}\ \cite{hpoc,Ellis}.

The NLO, order $\alpha_s^3$, corrections
to $Q \overline Q$ production have been calculated for the total cross section
\cite{NDE1}, single--inclusive quark distributions \cite{NDE2,SvN}, and
exclusive $Q \overline Q$ pair distributions \cite{MNR}.
In addition to real and virtual corrections to the LO
processes, quark--gluon scattering, $ q (\overline q) g \rightarrow Q
\overline Q  q(\overline q)$  is also included\footnote{This process
has been interpreted at LO as the scattering of a heavy quark
excited from the nucleon sea with a light quark or gluon, {\it e.g.}\ $gQ
\rightarrow gQ$, and is referred to as flavor excitation \cite{Ellis,NDE1}.
However,
for moderate $p_T$, flavor excitation is suppressed at LO due to the small
heavy quark parton density near threshold.}.
The total partonic cross section, $\widehat{\sigma}_{ij}$,
can be expressed as \be
\widehat{\sigma}_{ij}(x_1x_2s,m_Q,\mu) = \frac{\alpha_s^2(\mu)}{m_Q^2}
\left\{ f^0_{ij}(\rho) + \frac{\alpha_s(\mu)}{4\pi} \left[f^1_{ij}(\rho) +
\overline f^1_{ij}(\rho)\ln(\mu^2/m_Q^2) \right] + {\cal O}(\alpha_s^2)
\right\} \,\, , \ee where $\rho = 4m_Q^2/x_1x_2s$ and $f_{qg}^0 = 0$.
The double differential meson pair production cross section to NLO is then 
\be E_H E_{\overline H} \frac{d \sigma_{H \overline H}}{d^3p_H
d^3p_{\overline H}} & = & \sum_{i,j}
\int dx_1 dx_2 dz_H dz_{\overline H} \frac{E_H E_{\overline H}}{E_Q 
E_{\overline Q}} \frac{D_{Q/H}(z_H)D_{\overline Q/\overline H}(z_{\overline 
H})}{z_H^3 z_{\overline H}^3}
\\ \nonumber &   & \left[E_Q E_{\overline Q} \frac{d 
\widehat{\sigma}_{ij}(x_1x_2s,m_Q,\mu)}{d^3p_H d^3p_{\overline H}} \right]
q_i(x_1,\mu) q_j(x_2,\mu)  \, \, , \ee where $q_i$ and
$q_j$ are the quark, antiquark and gluon densities, appropriately defined in
the particular scheme.  The sum runs over $gg$ fusion, $q \overline q$
annihilation and $q(\overline q) g$ scattering.
The NLO corrections become large when $m_Q/\sqrt{s} \ll 1$
since gluon exchange dominates the
asymptotic behavior.  However, the perturbative expansion may still be
valid if further higher-order corrections are small.

The $c \overline c$ total cross section at NLO, $\sigma_{c \overline c}^{\rm
tot}$, has been compared with $pp$ and $pA$ data at $\sqrt{s} \leq 63$ GeV
\cite{Reu,MLM1}, assuming a linear
nuclear dependence \cite{E769,Pat,Appel}, to fix $m_c$ and $\mu$ and provide an
extrapolation to collider energies \cite{hpoc}.
Reasonable agreement with the data was
found for $m_c = 1.2$ GeV and $\mu=2m_c$ for MRS D$-^\prime$, leading to
$\sigma_{c \overline c}^{\rm tot} = 344 \, \mu$b at RHIC
and 17.7 mb at the LHC.  We use $m_b =
\mu = 4.75$ GeV to calculate $b \overline b$ production, finding
$\sigma_{b \overline b}^{\rm tot} = 1.5 \, \mu$b
at RHIC and 224 $\mu$b at the LHC. The $c
\overline c$ cross sections are larger than earlier estimates \cite{VJMR}
because of the low $x$ behavior of the gluon distributions\footnote{Choosing
another parton density with a lower initial scale but similar
small $x$ behavior \cite{GRV} leads to a
somewhat smaller charm cross section at LHC, $\sigma_{c \overline c}^{\rm
tot} = 6.7$ mb while the
$c \overline c$ cross section at RHIC is 351 $\mu$b, nearly independent of the
choice of parton densities \cite{hpoc}.}.  However, the
theoretical $K$ factor, $\sigma_{Q \overline Q}^{\rm NLO}/\sigma_{Q \overline
Q}^{\rm LO} \sim 2-3$, is rather large,
particularly for $c \overline c$, suggesting that the perturbative expansion is
unreliable for $m_Q/\sqrt{s} \ll 1$. For details of
the calculation and the theoretical uncertainties, see \cite{hpoc,rvsc}.

It was recently shown that when $\mu \propto m_T$, the theoretical
$K$ factor is nearly constant for distributions that are nontrivial at LO
\cite{rvsc}\footnote{Ref.\ \cite{Ina} reaches a different conclusion with $\mu
\propto m_Q$.  However, this choice introduces large logarithms for $p_T \geq
m_Q$, making the calculation unstable.}. Therefore, the NLO
calculation is essential only for the pair $p_T$ and azimuthal distributions.
We calculate the meson pair invariant mass distributions and the
double differential rapidity and rapidity
gap, $y_g = y_H - y_{\overline H}$, distributions $d\sigma_{H 
\overline H}/dydM$ and $d\sigma_{H \overline H}/dy_gdM$ at LO assuming 
$D_{Q/H}(z) =
\delta(1-z)$ and checked that our results agreed with the LO results from the 
program of Nason and collaborators \cite{MNR}. 
The rapidity gap is needed for acceptance studies since the mesons decay
independently. The total cross section and $Q \overline Q$ pair $p_T$
distributions are calculated at NLO using the same program
\cite{MNR}.  We note that, so far, no resummation of the $Q
\overline Q$ $p_T$ distribution
has been performed for low $p_T$, analogous to Drell-Yan production 
(see eq.\ (3) and \cite{hpdy,AK}).  
Therefore, we have only shown the $Q \overline Q$ $p_T$
distribution where the calculation may be considered to be reliable. 
The $Q \overline Q$ and $H \overline H$
$p_T$ distributions are assumed to be equal, in keeping with the trivial
fragmentation function used for the longitudinal momentum distribution 
\cite{VBH2}.
At large $p_T$, this is no longer a good approximation.  However, in the
low $p_T$ region, of interest in heavy-ion collisions, this assumption 
should be reasonable.
We assume no intrinsic light parton $p_T$ in the parton densities
and integrate over $x_1$ and $x_2$
using four-momentum conservation.  The triple differential cross section,
eq.\ (5) is then
\be \frac{d \sigma_{H \overline H}}{dy dy_g dM^2} & = & 
\frac{\sigma_{\rm NLO}}{\sigma_{\rm
LO}} \int dp_{T, Q}^2 dz_H dz_{\overline H} dy_H dy_{\overline H}
C(x_1, x_2) \frac{E_H E_{\overline H}}{E_Q E_{\overline Q}}
\frac{D_{Q/H}(z_H) D_{\overline Q/\overline H}(z_{\overline H})}{z_H 
z_{\overline H}} \nonumber \\
&   & \delta(y - (y_H + y_{\overline H})/2)
\delta(M^2 - (p_H + p_{\overline H})^2) \delta(y_g - y_H +
y_{\overline H}) \, \, , \ee where $\sigma_{\rm NLO}/\sigma_{\rm LO}$ is the
theoretical $K$ factor.  The definition of the pair rapidity, $y = (y_H +
y_{\overline H})/2$, holds when $p_{T, Q} = p_{T, \overline Q}$ which is 
true when the initial parton $p_T$ is neglected.

If the average number of $H \overline H$ pairs,
\be N_{H \overline H} = T_{AB}(0) \sigma_{H \overline H} \, \, , \ee
produced in central collisions is small,
$N_{H \overline H}  \ll 1$,
the lepton pairs will be correlated with $N_{ll}^{\rm corr} =
N_{H \overline H} B_R^2(H/\overline H \rightarrow l^\pm X)$.
However, if $N_{H \overline H}>1$, opposite sign lepton pairs
from uncorrelated
$H \overline H$ pair decays need to be taken into account.
When $N_{H \overline H} \gg 1$,
the average number of uncorrelated lepton pairs is
$N_{ll}^{\rm uncorr} = N_{H \overline H} (N_{H \overline H} -1)
B_R^2(H \overline H \rightarrow l^\pm X)$.
If $N_{H \overline H} \approx 1$, a distribution in $N_{H \overline H}$ must
be considered to calculate the uncorrelated pairs. 

The $c \overline c$ production cross sections are large enough for
lepton pair production from uncorrelated $D \overline D$ decays to be
substantial in nuclear collisions, particularly at the LHC.  Given our
values of $\sigma^{\rm tot}_{c \overline c}$ and assuming that all $c
\overline c$ pairs produce final-state $D \overline D$ pairs, we find
$N_{D \overline D} \sim 8.7$ at RHIC and $N_{D \overline D} \sim 450$
at the LHC\footnote{The $c \overline c$ production rate at LHC is
compatible with an earlier estimate using the parton cascade model
\cite{KG} where $\sim 400$ $c \overline c$ pairs are produced at
$\sqrt{s}=6.3$ TeV.  However, our RHIC rate is considerably smaller
than the 60 $c \overline c$ pairs predicted by the parton cascade.
The discrepancy cannot be fully accounted for within the theoretical
uncertainties.  In fact, if we use the same parton densities as in the
parton cascade model with the same charmed quark mass and scale, our
production rate drops by a factor of two at the LHC but is unchanged
at RHIC energies \cite{hpoc}.  Therefore, initial $c \overline c$
production cannot be the only important source of charm in the parton
cascade model.}.   Up to 3100 uncorrelated pairs can be formed
among the $B_R(D \rightarrow l^+ X)N_{D \overline D} \sim 55\ l^+$ and
$l^-$ leptons from the charm decays in central Pb+Pb collisions at the
LHC.

In Figs.\ 4 and 5 we show the rates of
$c \overline c$ and $D \overline D$ production at RHIC and LHC as a function
of $p_T$, $M$, $y_g$, and $y$.  The $b
\overline b$ and $B \overline B$ rates are given in Figs.\ 6 and 7.
The meson pair rapidity distributions are somewhat
narrower than the corresponding quark pair distributions due to the quark and
mason mass differences. The rapidity gap increases with $M$ and $dN/dy_gdM$
has a minimum at $y_g = 0$ since for fixed mass, $M^2 = 2m_T^2 (1 +
\cosh y_g)$, small $y_g$ corresponds to large $p_T$.
However, the mass integrated rapidity gap distribution, $dN/dy_g$, 
does not have a minimum at $y_g = 0$.

Lepton pairs from $D \overline D$ and $B \overline B$ decays
have been obtained using a Monte Carlo code based on
$D$ decays measured by Mark-III \cite{Balt} and $B$ decays observed with CLEO
\cite{CLEO}.
Opposite sign lepton pairs can be produced from a single $B$ decay by the chain
$B \rightarrow \overline D l^+ X$, $\overline D \rightarrow l^- X$.  Our
calculation assumes that the lepton pairs result from the decay of the initial
$B$ and $\overline B$.  Note that the CLEO analysis \cite{CLEO} 
assumes that the measured leptons are from the decays of the initial $B$'s 
rather than from secondary decays.  However, if $B \overline
B$ production is significant, the secondary decays in the chain could add a
component to the low mass background. 
The inclusive branching ratio for $D$ meson decay to leptons, averaged over
charged and neutral $D$'s, is $B_R(D^0/D^+ \rightarrow l^+ X) \sim 12$\%
\cite{PDG}.  The corresponding
branching ratio for $B$ mesons of unspecified charge is
$B_R(B \rightarrow l^+ X) \sim 12$\% \cite{PDG}.
The momentum vectors of each meson are computed
in the $H \overline H$ pair rest frame using the rapidity gap 
distribution, $dN/dy_gdM$, to separate the
mesons.  The decays are calculated in the meson rest frame according to the
measured lepton momentum distributions \cite{Balt,CLEO} and
boosted back to the nucleon-nucleon center of mass frame where the lepton pair
quantities are computed.  To account for uncorrelated lepton pair production in
the Monte Carlo code, two $H \overline H$ pairs are generated and the
$H$ from the first pair is decayed with the $\overline H$ from the second.

The lepton pairs from correlated and uncorrelated $D \overline D$
decays are shown for central Au+Au collisions at RHIC and Pb+Pb
collisions at LHC in Figs.\ 4 and 5.  The number of lepton pairs from
heavy quark decays at $M=2$, 4, and 6 GeV are given in Table 1 for
central Au+Au collisions at RHIC and in Table 2 for central Pb+Pb
collisions at the LHC.  Note that in obtaining the number of pairs
from our Monte Carlo simulation, we average over mass bins of width
400 MeV.  The average mass and $p_T$ of the decay pairs produced at
RHIC and LHC energies are shown in Table 3. The dileptons resulting
from uncorrelated $D \overline D$ pairs have larger masses since the
rapidity gap between the uncorrelated mesons is larger on average than
between the correlated pairs.  In fact the average mass of lepton
pairs from uncorrelated $D \overline D$ decays is 30\% larger than
that of the correlated pairs at LHC. The average rapidity of the
lepton pairs from uncorrelated $D \overline D$ decays is smaller than
from correlated decays at the LHC whereas the average $p_T$ is
somewhat larger for the uncorrelated pairs.  Uncorrelated $B \overline
B$ decays have not been considered here but should produce only a
relatively small enhancement at LHC for $M<10$ GeV.  We show the
lepton pairs from correlated $B \overline B$ decays in Figs.\ 6 and 7.
Note that, in general, the average lepton pair $p_T$ increases with
dilepton mass while the average lepton pair rapidity decreases with
mass.

\section{Dilepton Production by Thermal Processes}

We now calculate the rates of thermal lepton pair production, both
directly from thermal $q \overline q$ annihilation and from the decays of
$D \overline D$ pairs produced thermally. 
We expect that the longitudinal expansion approximately follows
the scaling law,
$v_z=z/t$, at RHIC and LHC until large rapidities where particle
densities become small and pressure gradients cause scaling violations.
Numerical calculations \cite{KaRaRu,Kat2} give support for a picture where
these deviations are small up to the fragmentation regions, therefore we neglect
them here.

With the scaling ansatz for the longitudinal velocity we can relate the
initial density and initial time to the final multiplicity
distribution of hadrons
by $n_i \tau_i = (dN/d\eta)/(\pi R_A^2)$.  The temperature is therefore a
function of $\eta$, $T_i \equiv T_i(\eta)$,
where we can identify the fluid and particle rapidities since
the overall rapidity distribution is much broader than
the thermal distribution at freeze-out.
We parametrize the multiplicity distribution in the form
\be \frac{dN}{d\eta} =
\left(\frac{dN}{d\eta}\right)_0 \exp(-\eta^2/2\sigma^2) \, \, , \ee
where $(dN/d\eta)_0$ is the total multiplicity at $\eta =0$
and $\sigma$ measures the width of the hadron rapidity distribution.  As 
before, \cite{VJMR} we assume the following initial conditions:
at RHIC energies, $|\eta| \leq 6$ and we choose $(dN/d\eta)_0 = 2000$
while for the LHC $|\eta| \leq 8$ and we take $(dN/d\eta)_0 = 5000$.  The
width is estimated from the total energy of the final particles.
We find $\sigma\sim~3$ at RHIC and at LHC, $\sigma \sim 5$.  These 
rapidity densities, while somewhat high, are consistent with the (optimistic)
high initial temperatures taken here, as well as with Parton Cascade Model 
calculations \cite{KG}.  Like the scaling properties of the
longitudinal velocity, the Gaussian form of the multiplicity distribution must
break down at rapidities close to the phase space limit. There, however, the
densities become too small for thermal production to be important.

\subsection{Thermal Dileptons}

In the quark-gluon plasma, lepton pairs are expected to be produced by the
annihilation process $q
\overline q \rightarrow \gamma^\star \rightarrow l^+ l^-$, similar to 
Drell-Yan production in the initial nucleon-nucleon interactions.  However, in
the plasma, the nuclear quark distributions are replaced by thermal
distributions.
Neglecting the transverse flow the dilepton emission rate is then \cite{KR,OP}
\be \frac{dN}{d^4x d^2p_T dy dM^2} = \frac{\alpha^2}{8\pi^4} F \exp{ [-(m_T
\cosh(y - \eta)/T)]} \, \, , \ee where $T$ is the temperature and $\eta$
is the rapidity of the medium. In a plasma, $F_Q = \sum
e_q^2$, while in an equilibrium hadron gas, assuming
$\pi \pi \rightarrow \rho \rightarrow l^+ l^-$
is the dominant channel, $F_H = \frac{1}{12} m_\rho^4 /((m_\rho^2 -
M^2)^2 + m_\rho^2 \Gamma_\rho^2)$.
The mixed phase has a fractional contribution from each with $F_M = f_0
(r-1) \left[f_0 F_Q + ((r-2)f_0 + 2)F_H \right]$.

In addition to $\pi \pi \rightarrow \rho \rightarrow l^+ l^-$ other
hadronic processes contribute to the dilepton rate. In the region above the
$\rho$ mass they dominate \cite{GaLi} over the $\rho$ contribution which is
cut off by the form factor. The assumption that vector mesons saturate the
$e^+e^- \to {\rm hadrons}$ cross section indicates that at the phase transition
temperature the dilepton rate in the hadron gas could be very similar to that
from the quark-gluon plasma \cite{LeRu}. Neglecting these contributions
underestimates the hadronic
emission in the mass range between the $\rho$ and $J/\psi$ mesons.

The rapidity distribution for a given dilepton mass is
\be \frac{dN}{dM dy} & = & \frac{\alpha^2}{2 \pi^3} \pi R_A^2 \left\{ \theta
(s_i - s_Q) \frac{3 F_Q }{M^3} \int_{-Y}^{Y}
\frac{d\eta (\tau_i T_i^3)^2}{\cosh^6(\eta - y)} P_1(x)
e^{-x}|_{x_c}^{x_i} \right. \nonumber \\
&~& \mbox{} + \theta(s_i - s_H) M^3 \frac{\tau_m^2}{2} F_M \int_{-Y}^{Y} d\eta
\left( \frac{1}{x_c} + \frac{1}{x_c^2} \right)e^{-x_c} \\
&~& \mbox{} + \left. \theta(s_i - s_{\rm dec}) \frac{3 F_H}{M^3} \int_{-Y}^{Y}
\frac{d\eta (\tau_H T_H^3)^2}{\cosh^6(\eta - y)} P_1(x)
e^{-x}|_{x_{\rm dec}}^{x_H} \right\} \, \, , \nonumber
\ee where $P_1(x) = x^4 + 5x^3 + 15x^2 + 30x + 30$,
$x = M \cosh(\eta - y)/T$, $s_i$ is the  initial entropy density and
$s_Q$ and $s_H$ are the entropy densities of the  plasma and the hadron gas at
the transition temperature $T_c$.
The invariant mass distribution is obtained by integrating eq.\ (13) over
the rapidity interval $-Y \leq y \leq Y$.
Likewise the transverse momentum distribution for mass $M$ is
\be \frac{dN}{dM dp_T} & = & \frac{\alpha^2}{2 \pi^3} \pi R_A^2 M p_T
\left\{ \theta (s_i - s_Q) 3 F_Q  \int_{-Y}^{Y}
\frac{d\eta dy (\tau_i T_i^3)^2}{m_T^6 \cosh^6(\eta - y)} P_2(z)
e^{-z}|_{z_c}^{z_i} \right. \nonumber \\
&~& \mbox{} + \theta(s_i - s_H) \frac{\tau_m^2}{2} F_M
\int_{-Y}^{Y} d\eta dy e^{-m_T \cosh(\eta - y)/T} \\
&~& \mbox{} + \left. \theta(s_i - s_{\rm dec}) 3 F_H \int_{-Y}^{Y}
\frac{d\eta dy (\tau_H T_H^3)^2}{m_T^6 \cosh^6(\eta - y)}  P_2(z)
e^{-z}|_{z_{\rm dec}}^{z_H} \right\} \, \, , \nonumber
\ee where $P_2(z) = z^5 + 5z^4 + 20z^3 + 60z^2 + 120z + 120$ and $z = m_T
\cosh(\eta - y)/T$.

As defaults, we assume a three-flavor plasma and a hadron gas of massless
pions with $T_c = 200$ MeV and $T_{\rm dec} = 140$ MeV.  Then
\[ s_i = \left\{
\begin{array}{ll} 4 \gamma_k \displaystyle{ \frac{\pi^2}{90}}T_i^3 & 
\mbox{$s_i > s_Q$ or $s_i
< s_H$} \\ f_0 s_Q + (1-f_0)s_H & s_H < s_i < s_Q \end{array} \right. \]
where $\gamma_k$ is the number of degrees of freedom with
$\gamma_H = 3$ and $\gamma_Q = 16 + 21n_f/2$ and $n_i = s_i/3.6$.
The beginning of the mixed phase occurs at \[
\tau_m = \left\{ \begin{array}{ll} \tau_i (T_i/T_c)^3 & T_i > T_c \\ 1 \, {\rm
fm} & T_i = T_c \, \, . \end{array} \right. \]  The initial plasma content of
the mixed phase is \[ f_0 = \left\{ \begin{array}{ll} 1 & T_i > T_c \\ (s_i
- s_H)/(s_Q - s_H) & T_i = T_c \, \, . \\ 0 & T_i < T_c \end{array} \right. \]
The beginning of the hadron phase is then \[
\tau_H = \left\{ \begin{array}{ll} \tau_m r & T_i \geq T_c \, \, ,
\\ 1 \, {\rm fm} & T_i < T_c  \end{array} \right. \]
where $r = s_Q/s_H = \gamma_Q/\gamma_H$.
There is no thermal contribution if $T_i < T_{\rm dec}$.

In our previous work, we used three different hypotheses to fix $T_i$
and $\tau_i$ \cite{VJMR}.  Since here we wish to determine if any
plasma contribution is observable at all, we focus on the highest
temperature scenario, $\tau_i (3 T_i) \simeq \hbar c$, yielding
$T_{i,{\rm max}} \sim 515$ at RHIC and 810 MeV at the LHC, similar to
the results of Ref.\ \cite{Joe}.  The corresponding initial times are
$\tau_i \sim$ 0.2 fm and 0.08 fm respectively.  Thermalization at such
early times is a bold assumption, if taken literally.  Following
McLerran, Kapusta and Srivastava \cite{Joe}, we employ it to
schematically characterize pre-equilibrium pair production
\cite{GeKa,AlSe}.  For our choice of initial conditions, plasma
production dominates the thermal distributions over all observable
phase space. However, other initial conditions with lower initial
temperatures result in a much smaller thermal yield for $M \geq 2$
GeV. For example, if we choose instead $T_c = 150$ MeV and $T_{\rm
dec} = 100$ MeV, as recent lattice calculations with quarks indicate
\cite{Karsch95}, then the hadron gas contribution is considerably
reduced.  Since we have assumed a high initial temperature and study
pairs with $M\geq 2$ GeV only, the hadron contribution is a small
fraction of the yield.  At $M=2$ GeV, the yield is reduced by
$\sim$30\% with $T_c = 150$ MeV while higher masses are virtually
unaffected.  Reducing $(dN/d\eta)_0$ would reduce $T_i$, increasing
the relative importance of the hadronic contribution and decreasing
the plasma yield.  Additionally, recent analyses of the initial
conditions \cite{KariQM,EKR} suggest that the quark density is too low
for the quarks to be in chemical equilibrium.  Therefore, even if the
initial temperature is high, the dilepton yield would be significantly
reduced because of the low quark density.

We neglect the transverse expansion and describe the matter
in the longitudinal direction only.  Transverse expansion mainly affects
the later part of the mixed phase and the hadron gas phase \cite{PVR,Kat1}.
If it is included, the contribution from the hadron gas
phase is negligible for $M\geq 2$ GeV except at large rapidities where the
initial density is too low to produce a plasma.  Therefore a
full three-dimensional hydrodynamic
calculation would give a narrower thermal pair rapidity distribution than 
our longitudinal scaling approximation.

We have so far discussed the properties of the lepton pairs.  However,
once acceptance questions are addressed, it is necessary to track each
lepton separately.
The space-time integration remains the same, but now we write
more generally
\be 
E \frac{dN}{d^4x d^3p dy_g dM^2} & = & \int \frac{d^3k_1}{E_{k_1}}
\frac{d^3k_2}{E_{k_2}} \frac{d^3p_1}{E_{p_1}} \frac{d^3p_2}{E_{p_2}} f(k_1)
f(k_2) \frac{36|{\cal M}|^2}{16(2 \pi)^8}  \delta^4(k_1 + k_2 - p_1 - p_2)
\nonumber \\
&~& \mbox{}  \delta(M^2 - (p_1 + p_2)^2) \delta(y_g - y_{p_1} +
y_{p_2}) \delta(\vec p - \vec p_1 - \vec p_2) \, \, , \ee
where $k_1$ and $k_2$ are the incoming quark and antiquark, $p_1$ and $p_2$
are the outgoing leptons, and $f(k)$ is the thermal (Boltzmann)
distribution of the initial partons.  Note that if the rapidity
gap is integrated over, eq.\ (15) reduces to eq.\ (12).  The $p_T$, mass,
rapidity gap, and pair rapidity
distributions for $M \geq 2$ GeV are shown in Figs.\ 8 and 9 for
RHIC and LHC respectively.  The number of thermal lepton pairs with $M=2$, 4,
and 6 GeV are given in Tables 1 and 2 for central collisions at RHIC and LHC.
The rapidity gap distribution, $dN/dy_gdM$ does not increase with $y_g$ 
at fixed $M$ as it does
for the initial $Q \overline Q$ production because the thermal partons are
not aligned along a particular direction.  However, the average $p_T$ and
rapidity of the thermal distributions follows the trend of the heavy quark
decays: $\langle p_T \rangle$ increases with mass while $\langle y \rangle$
decreases.  The average thermal pair mass and $p_T$ are shown in Table 3 for
$M>2$ GeV at RHIC and LHC.

\subsection{ Thermal Charm}

With a large initial temperature in the most optimistic scenario,
$T_{i,{\rm max}} \sim (1/3 - 1/2) m_c$,
significant thermal charm
production may be expected, as previously suggested \cite{Shuryak,Shor}.
The thermal $c \overline c$ production rate can be found by making the
replacement
\be 36|{\cal M}_{q \overline q \rightarrow ll}|^2
\rightarrow \gamma_q |{\cal M}_{q \overline q \rightarrow c
\overline c}|^2 + \gamma_g |{\cal M}_{gg \rightarrow c \overline c}|^2
\, \, , \ee in eq.\ (15)
where now $p_1$ and $p_2$ are the four-momenta of the charm quarks and
the quark and gluon degeneracy factors are $\gamma_q = 3 \times (2 \times
3)^2$ for three quark flavors and $\gamma_g = (2 \times 8)^2/2$ (the factor of
1/2 is needed to prevent double counting) \cite{Shor}.
The matrix elements are given
in \cite{Comb}.  We use $m_c = 1.2$ GeV to be consistent with 
the initial $c \overline c$
production calculations and include only plasma production\footnote{In our
previous paper \cite{VJMR}, 
a normalization constant was left out of the calculation.  The
correction has the effect of reducing the thermal charm yield.  Note 
also that a typographical 
factor of two is also missing from eq.\ (13) of that paper.}.
The fragmentation of the charmed quarks into $D \overline D$ pairs has also
been incorporated, as in eq.\ (5).  
We find approximately one thermal $D \overline D$ pair
at RHIC and 23 pairs in Pb+Pb collisions at LHC.  These
results are essentially in agreement with those of the ideal thermal case
discussed in \cite{LMW} and \cite{LG}.  The rate is high enough at
the LHC for $\approx 500$ lepton pairs to be produced from uncorrelated decays.
Again, an ideal background subtraction should remove the uncorrelated pairs.

The thermal charm and resulting $D \overline D$ distributions are
shown in Figs.\ 10 and 11 along with the dilepton yields.  The number
of thermal lepton pairs with $M=2$, 4, and 6 GeV are given in Tables 1
and 2 for central collisions at RHIC and LHC.  As before, the number of
pairs at each mass is averaged over mass bins of width 400 MeV.  The
average $D \overline D$ pair mass and $p_T$ are somewhat larger than
the corresponding $c \overline c$ averages.  The $D \overline D$ pair
mass is 1.5 GeV larger while $\langle p_T \rangle$ is 400 MeV larger
at RHIC and 600 MeV larger at LHC.  These increases deplete high
rapidity $D \overline D$ pairs with respect to $c \overline c$
production.  The rapidity gap is also decreased, although not
significantly.  The average thermal pair mass and $p_T$ are given in
Table 3 for both RHIC and LHC.  The mass and $p_T$ distributions of
lepton pairs resulting from correlated $D \overline D$ decays are
similar in shape to the thermal dileptons and with a similar yield at
the same lepton pair mass.  However, the lepton pair rapidity
distributions from $D \overline D$ decays are narrower.  As in the
case of initial $c \overline c$ production, the uncorrelated lepton
pairs have a much broader mass distribution than the correlated pairs.

\section{Results and Conclusions}
\subsection{Results}

We now compare the perturbative lepton pair production with thermal
production.  The mass distributions have already been shown in Fig.\ 1
for RHIC and LHC although without the uncorrelated $D \overline D$
decays.  The $p_T$ and rapidity distributions for lepton pairs with
$M=2$, 4, and 6 GeV are shown in Fig.\ 12 for RHIC and 13 for LHC.
Here the yield from the
uncorrelated $D \overline D$ decays are included.  The average lepton
pair masses for $M>2$ GeV and transverse momenta at both energies are
given in Table 3.  The average $p_T$ of the lepton pairs are given for
$M>2$ GeV only for Drell-Yan and thermal dilepton production.  The
average $p_T$ from heavy quark decays, initial and thermal, are given
for all $M$.

Uncorrelated initial $D \overline D$ decays dominate the distributions,
by several orders of magnitude 
at the LHC.  However, the like-sign subtraction that removes
$\pi^+ \pi^-$ and $K \overline K$ decays from the continuum should also remove
the uncorrelated $D \overline D$ decays.  The correlated $D \overline D$ decays
are part of the signal and will not be subtracted.  These decays dominate the
continuum up to $M=10$ GeV.  At higher masses, the $B \overline B$
decays begin to be as important as correlated $D \overline D$ decays.
The same results are observable in the
$p_T$ and rapidity distributions.  At RHIC, the contributions from
the initial hard processes are above the thermal contributions
over all phase space except at $M=2$ GeV and $y>4.5$, as seen in Fig.\ 12(b).
The thermal
dilepton and thermal charm contributions are somewhat above those of the 
$B \overline B$
and Drell-Yan pairs for $M<3$ GeV at RHIC.   Both thermal contributions have 
very similar distributions and yields at
RHIC energies.  The thermal dilepton and thermal $D \overline D$
distributions are also similar 
at the LHC although the thermal charm yield is larger than the thermal 
lepton pair yield.  At the LHC, the $B \overline B$
decays produce more lepton pairs than the thermals for $M>2$ GeV.  However,
the correlated thermal charm yield is above the $B \overline B$ decays at $M=2$
GeV. The uncorrelated thermal charm yield is larger than the $B \overline B$
decay rate.  If all the heavy quark decays could be subtracted,
there might be a small window
of opportunity to observe thermal dileptons, both prompt and from charm decays,
over Drell-Yan production at low $p_T$, as seen in Fig.\ 12.  Because RHIC
is at a significantly lower energy than the LHC, the slopes of the rapidity
and $p_T$ distributions of each of the contributions are somewhat different,
even in the central region, as seen in Fig.\ 12.
The difference in the slopes could perhaps help disentangle
the dilepton sources, if very large rapidities could be measured.  
All the LHC $p_T$ and rapidity spectra have similar slopes, see Fig.\ 13,
making differentiation more difficult.

Our results suggest that it is very unlikely that the thermal $q \overline
q$ annihilation signal can be
extracted. However, the experimental acceptance has not been included.  The
acceptance should be smaller
for lepton pairs from $D \overline D$ decays than for Drell-Yan or thermal
dileptons.
Since, especially for $D \overline D$ decays, 
large lepton pair mass implies a large rapidity gap,
at least one of the $D \overline D$ decay leptons may be outside the 
finite detector acceptance.
In particular, relatively few high mass lepton pairs from uncorrelated
$D \overline D$ decays will be detected, significantly reducing the
uncorrelated yield even before like-sign subtraction.
The $B \overline B$ decay pairs will have
a larger acceptance due to the increase in pair $p_T$ over
the $D \overline D \rightarrow l^+ l^-$ decays at the same mass which reduces
the rapidity gap.
Therefore a judicious choice of kinematic cuts can reduce the initial $D
\overline D$ acceptance relative to other dilepton sources that
produce nearly equal but opposite transverse momentum leptons with $p_T<M$.
To show how the finite detector acceptance changes the yield, we choose some
realistic cases to examine in more detail.

\subsection{RHIC: PHENIX}

The RHIC PHENIX detector is specially designed to measure
electromagnetic probes.  It consists of two central electron arms with
a rather small acceptance, $|\eta| \leq 0.35$ and $\pm 30^\circ < \phi
< \pm 120^\circ$, one muon arm in the forward direction \cite{CDR}
covering the pseudorapidity region $1.1 \leq \eta \leq 2.4$, and a
second muon arm in the backward direction with a similar angular
coverage \cite{PHspin}.  We assume that electrons with momentum
greater than 1 GeV and muons with momentum greater than 2 GeV can be
efficiently detected.  Although we choose a rather high momentum
cut-off, lower momentum leptons will be measured as well.
Additional coverage from the electron arms together with the muon arm
will partially fill in the rapidity gap between the two detector
systems through electron-muon coincidence studies.  The rapidity gap
between the lepton pairs becomes particularly important when finite
acceptance cuts are applied.  Pairs with a small rapidity gap are more
likely to be accepted, particularly in the central electron arms.

The detector acceptance can have a substantial effect on the relative rates, as
shown in Fig.\ 14 for (a) the central electron arms, (b) the forward muon arm,
and (c) the combined $e \mu$ coverage for PHENIX.  Note that some of
the distributions are limited by statistics.  For each system, the
accepted lepton pair mass distributions are shown.  The percentage of
accepted pairs with $M>2$ GeV and the average lepton pair mass
from all our sources in this mass range are given in Table 4.  In principle all
finite masses are accepted, as suggested from the $H \overline H$ decay
distributions in Figs.\ 4-7.  
However, since the Drell-Yan and thermal dileptons calculations
are most reliable for $M> 2$ GeV, we use $M=2$ GeV as a lower bound on the
accepted masses.  Note that
many of the $D \overline D$ decay pairs, both correlated and uncorrelated, have
an invariant mass less than 2 GeV so that the additional mass 
cut to compare the
number of accepted pairs on an equal footing significantly reduces the total
acceptance, especially for the uncorrelated $D \overline D$ pairs.

The finite acceptance strongly reduces the uncorrelated $D \overline D$ 
production relative to the correlated production, as seen in Fig.\ 14.  In the
central detector, $D \overline D$ decays have the smallest acceptance due to
the relatively large rapidity gap between the leptons.
Particularly, the high mass uncorrelated pairs are removed from the spectrum.  
The $B \overline B$ decays have the largest acceptance here because the
combination of the relatively small pair rapidity and rapidity gap favors their
detection.  They will have the largest contribution to the continuum for $M>6$
GeV, after the correlated $D \overline D$ signal is negligible.

In the forward muon arm, the Drell-Yan and $B \overline B$ decays 
have very similar yields for
$M>3$ GeV since the $B \overline B$ acceptance will be decreased
relative to the tightly correlated Drell-Yan and thermal production with
their broader rapidity distributions. Although both correlated and 
uncorrelated $D \overline D$ decays are reduced
relative to the smaller rapidity gap of the $B \overline B$ decays,
the uncorrelated $D \overline D$ acceptance is reduced still
further because of the
larger rapidity gap between the uncorrelated lepton pairs.  In general, the 
broader rapidity coverage in the muon arm increases the acceptance of both the
correlated and uncorrelated $D \overline D$ pairs.  After like-sign
subtraction, the initial $D \overline D$ decays will dominate the spectrum
for most of the pair masses studied. 

In Fig.\ 14(c) we show only the acceptance for correlated and uncorrelated 
initial
$D \overline D$ decays to $e \mu$ pairs.  The increased rapidity coverage of
the combined system results in the acceptance of more low mass uncorrelated
$D \overline D$ decays.  We have not included the yield from any of
the other sources since $e \mu$ pairs cannot be produced by correlated
Drell-Yan and thermal production.  The rate must be large enough for at least
two lepton pairs to be produced per event for uncorrelated $e \mu$ pairs to
be important (except for $B \overline B$ decays) and the yields
from the other sources are small enough for such production to be unlikely (see
Table 1).

If the uncorrelated $D \overline D$ decays can be completely removed
by a like-sign subtraction, then $e \mu$ coincidence is a good way to
extract the correlated $D \overline D$ yield, which constitutes the
charm signature.  Even with a complete charm measurement, thermal
sources will be hard to detect, especially the thermal dileptons.  In
the muon arm, Drell-Yan and $B \overline B$ decays have nearly the
same rate and could be hard to separate.  However, in the electron
arms, for $6<M<10$ GeV, $B \overline B$ decays are most important and
could be removed from the lower mass continuum by a comparison of the
slopes.  The thermal $D \overline D$ signature may be measured below 3
GeV if the initial charm production can be reliably subtracted.  At 2
GeV in the central detectors, the accepted thermal $D \overline D$
yield is five times larger than the $B \overline B$ yield.  At low
mass and $p_T$, thermal charm could also be seen in the pair $p_T$
distributions.  Measurements from all the detector systems must
complement each other for reliable results.

\subsection{ LHC: ALICE}

At the LHC, only one heavy-ion detector, ALICE, is planned.  It includes a 
central detector that will measure dielectrons with
$M \leq M_{J/\psi}$ \cite{Alice} and covering $|\eta| \leq 0.9$.  We select 
leptons with momentum greater than 1 GeV. The ALICE collaboration has also
proposed a forward muon spectrometer, with $2.4 \leq \eta \leq 4$ to cover
higher mass pairs \cite{Alice2}.  We
consider only muons with momentum larger than 4 GeV in the muon arm.

The mass distributions for (a) the central detector and (b) the forward muon
spectrometer are shown in Fig.\ 15.  The relative rates
are similar to the corresponding RHIC detector systems although, overall, the
acceptances are larger in the ALICE detector, as shown in Table 5.
In the central detector, this is probably due to the full azimuthal
coverage and the larger rapidity coverage.
While the acceptance cuts substantially reduce the rate from uncorrelated $D
\overline D$ decays compared to the correlated decays,
the uncorrelated yield is still nearly an order of magnitude larger
than the correlated yield for masses below 4 GeV.  This higher acceptance for
uncorrelated pairs means than an accurate like-sign subtraction is crucial.
An additional background comes from uncorrelated thermal $D \overline D$
decays.  In the central detector, the uncorrelated thermal $D \overline D$ rate
is as large as the initial correlated $D \overline D$ rate at low masses.
However, in the muon arm, the $B \overline B$ decays are clearly the most
important source of lepton pairs after the initial charm production
due in part to the smaller rapidity gap between the
leptons---it is more likely that both decay leptons will lie within the
rapidity window of the detector.  The acceptance does not decrease in the muon
arm because the pair rapidity distribution is not significantly reduced with
respect to the Drell-Yan and thermal pair distributions, as is the case at
RHIC. The correlated
thermal $D \overline D$ yield may also be observable at $M \approx 2$ GeV
although the $B \overline B$ decay rate is within a factor of two here and more
difficult to remove reliably.  
The average mass of the accepted pairs in ALICE are also
given in Table 5.  The observed trends are similar to the PHENIX results except
that the conclusion that the thermal dilepton signal is unlikely to be 
measured is even stronger here although thermal charm may still be observable.

\subsection{Conclusions}

Charm production is the dominant source of dileptons in
heavy-ion collisions, even with acceptance cuts, for $M< 6-8$ GeV.
Uncertainties in QCD calculations may change the rates by a factor of two at
RHIC and 3-4 at LHC, not enough to affect this conclusion.  
Charm is both signal and background because the multiple
$c \overline c$ pair production results in substantial
uncorrelated $D \overline D$ contributions to the background.  We have only
included heavy quark production by first collisions.  However, multiple hard
scattering in $AB$ collisions can increase the charm yield 
before equilibration.

Nuclear shadowing is not yet well understood for the gluon.  
If the shadowing effects can be mapped out in phase space, the detection of 
thermal signals could be improved.  
Since $pA$ studies are planned at RHIC, nuclear shadowing could be mapped out
in phase space.  Such measurements are themselves important results.
The effects are strongest at the LHC where the saturation of the shadowing 
curve is reached.  Unfortunately this saturation region is
unlikely to be probed at RHIC and no corresponding $pA$ measurements can be
performed at the LHC so that the shadowing effects may be more difficult
to interpret.  It is clear that systematic
studies of charm production in $pp$, $pA$ and $AB$ interactions at the same
energy are needed to fully understand charm production.

We stress that our work differers from previous efforts \cite{Joe, 
GeKa} primarily in our  estimate  of  the perturbative background.  
Kapusta, McLerran and Srivastava \cite{Joe} assumed
hydrodynamic initial conditions similar to ours, but concluded that thermal
dileptons dominate the continuum below  the $\Upsilon$.
We attribute this striking difference to their estimate of the initial
hard scattering processes.  First, they omitted the contribution from 
semileptonic  charm decays.  They also underestimated the Drell-Yan 
contribution by
using  Duke-Owens parton distributions \cite{DO}, long obsolete.  
This same Drell-Yan estimate was also used in the comparison with dilepton
production by the parton cascade model \cite{GeKa}.  
 
Our calculations show that charm cannot be omitted.  The 
production of charm-decay and Drell-Yan dileptons  in primary 
collisions in any parton cascade or hydrodynamic model must agree 
with perturbative QCD and, therefore,  the rates should be similar to ours.  
Other initial charm calculations \cite{LMW} are within a factor of two to three
of ours, within the theoretical uncertainties involving the quark mass and
scale.  The $c \overline c$ production rate in the parton cascade model agrees
with perturbative QCD in $pp$ collisions but overestimates the importance
of flavor excitation \cite{Ellis}, leading to a larger charm yield than ours,
particularly at RHIC \cite{KG}.  However, we have omitted cascading of the 
$c$ and $\overline c$  quarks in the high  density medium, which can lead to
energy loss \cite{GM} (similar to `jet quenching' \cite{Miklos}).  If 
this loss is sufficient, these quarks can be equilibrated with the 
flowing plasma.  Since it is highly unlikely that all of the 
$c\overline c$ pairs can annihilate,  cascading will not change the number 
of pairs appreciably.  

We expect that thermal charm will  prove to be an experimentally accessible
temperature probe at RHIC and LHC.  On the other hand,  we  emphasize that 
thermal $q \overline q$ annihilation,  perhaps the  more familiar  thermal 
signal, will be much more difficult to pick out.  Thermal annihilation 
would be a more direct thermometer because the kinematics of the lepton
pair specifies the off-shell photon's four momentum.   However,  the heavy 
quark and Drell-Yan contributions are too high for the steeply-falling thermal
contributions to be extracted, unless the charm contributions can be reliably 
subtracted.

What is the best way to measure charm?  Coincidence measurements of $e
\mu$ can prove useful.  Charm was first measured by this method at the
ISR \cite{Chil} and such coincidence measurements are planned for
PHENIX \cite{CDR}.  Pairs of like-sign electrons may also offer a
measure of uncorrelated charm production.  Charm was measured with
single electrons by a study of the $e/\pi$ ratio at the ISR
\cite{Mike}.  Additionally, semileptonic decays can be experimentally
tagged and separated from direct production of lepton pairs with a
vertex detector. If a detached vertex is observed for at least one of
the leptons, then direct production of the pair can be ruled out. Such
a vertex detector is planned for the STAR detector \cite{STAR} at
RHIC.  Another technique for reducing the signal levels from
semileptonic decays is the use of selective kinematic cuts. Since the
leptons from decays have a weaker correlation in rapidity or angle
than those from directly produced pairs, cuts can be placed on these
variables. While some signal events will be lost, the signal to
background ratio can be improved for large acceptance detectors.

SG, PVR, and RV thank the Institute for Nuclear Theory at the
University of Washington in Seattle for their hospitality.  RV thanks
Brookhaven National Laboratory and the University of Jyv\"{a}skyl\"{a}
for their hospitality.  We are grateful to J. Carroll, K. Geiger,
W.~M. Geist, B. Jacak, D. Jouan, V. Koch, J.  Smith and E.~V. Shuryak
for discussions and K. J. Eskola for providing the shadowing
parameterizations used in the paper.

\appendix
\section{Appendix:  Shadowing Effects on Initial Processes}

\setcounter{equation}{0}

When the charged parton distributions are probed in deep-inelastic scattering
with a nuclear target and compared to a deuterium target, the ratio $R_{F_2} =
F_2^A/F_2^D$
has a characteristic shape as a function of $x$.  The low $x$ region,  
below $x\sim 0.1$, is referred to as the shadowing region, and between $x\sim 
0.3$ and 0.7 is the EMC region.  In both regions a depletion is observed in
the heavy nucleus relative to deuterium.  At very low $x$, $R_{F_2}$ appears to
saturate \cite{E6652}.  Between the shadowing and EMC 
regions, an enhancement occurs, called antishadowing, where
$R_{F_2}>1$.  There is also an enhancement as $x \rightarrow 1$, assumed to be
due to Fermi motion of the nucleons.  The entire nuclear dependence is often
referred to as shadowing.  Although the behavior of $R_{F_2}$ is 
not well understood, the effect has been described by
either an interplay of coherent and incoherent multiple scatterings in the
target or a modification of the parton densities in nuclear matter.  In any
case, the effect can be modeled by an $A$ dependent fit to the nuclear
deep-inelastic scattering data and implemented by a modification of the parton
distributions in the proton.  
In this appendix, we show the effect of two different
parameterizations of the nuclear parton densities to illustrate how the
Drell-Yan and $Q \overline Q$ distributions calculated for $pp$ interactions
might change in nuclear collisions at RHIC and LHC.  

In the central region at RHIC and LHC, the values of $x$ probed are
small enough for the hard processes we consider to be predominantly in
the shadowing region.  However, the momentum fractions increase with
pair mass, transverse momentum, and rapidity.  At $y=0$ and $p_T = 0$,
$x \sim M/\sqrt{s}$ so that in the mass range $2<M<6$ GeV, $0.01 < x <
0.03$ at RHIC and $3.6 \times 10^{-4} < x < 1.09 \times 10^{-3}$ at
the LHC.  It follows that $R_{F_2}$ may change significantly
at RHIC.  At the LHC, $x$ is small enough for the shadowing to be
saturated over most measurable rapidities.  The most important point
to note is that the reduction in the Au+Au cross section is never
significantly more than a factor of two for any of the processes
considered.  If the shadowing function can be mapped out in $pA$
interactions at RHIC, as discussed in Ref.\ \cite{LG2}, then the
corrections to $AA$ interactions may be relatively clear for $Q
\overline Q$ and Drell-Yan production, especially since the lepton
pairs from $Q \overline Q$ decays should reflect the shape of the
shadowing function in the same way as the $Q \overline Q$ pairs
themselves \cite{LG2}.

The first parameterization is a general fit to the most recent nuclear 
deep-inelastic scattering
data.  The fit does not differentiate between quark, antiquark, and gluon
modifications and does not include evolution in $Q^2$.
It is not designed to satisfy the baryon number or momentum sum rules.
The functional form of $R_{F_2}$ is \cite{EQC} \be R_{F_2} = \left\{
\begin{array}{ll} R_s {\displaystyle \frac{1 + 0.0134 (1/x -1/x_{\rm sh})}{1
+ 0.0127A^{0.1} (1/x - 1/x_{\rm sh})}} & \mbox{$x<x_{\rm sh}$} \\
a_{\rm emc} - b_{\rm emc}x  & \mbox{$x_{\rm sh} <x< x_{\rm fermi}$} \\
R_f \bigg( {\displaystyle \frac{1-x_{\rm fermi}}{1-x}} \bigg)^{0.321} &
\mbox{$x_{\rm fermi} <x< 1$} \end{array} \right. \, \, , \ee
where $R_s = a_{\rm emc} - b_{\rm emc} x_{\rm sh}$, $R_f = a_{\rm emc} -
b_{\rm emc} x_{\rm fermi}$, $b_{\rm emc} = 0.525(1 - A^{-1/3} - 1.145A^{-2/3} +
0.93A^{-1} + 0.88A^{-4/3} - 0.59A^{-5/3})$, and $a_{\rm emc} = 1 + b_{\rm emc}
x_{\rm emc}$.   The fit fixes $x_{\rm sh}=0.15$, 
$x_{\rm emc}=0.275$ and $x_{\rm fermi}=0.742$.
In Fig.\ 16(a) we show $R_{F_2}(x)$ for $A=197$.  
In the nucleus, the parton densities are  modified so
that \be q_f^A(x,\mu) & = & R_{F_2}(x) q_f^p(x,\mu) \\ g^A(x,\mu) & =
& R_{F_2}(x) g^p(x,\mu) \, \, , \ee where $q_f$ represents both valence and sea
quarks.  Since $R_{F_2}(x)$ is scale independent and the parton densities are
treated equivalently, the ratio of hard process production in Au+Au
to $pp$ collisions at the same energy is \be R(y,p_T,M) = \frac{d\sigma_{\rm
AuAu}/dydp_TdM}{d\sigma_{pp}/dydp_TdM} \propto R_{F_2}(x_1)
R_{F_2} (x_2) \, \, . \ee

In Fig.\ 16(b)-(d) and Fig.\ 17 we show the nuclear effect on
leading order calculations of heavy quark and Drell-Yan production.
Figures 16(b)-(d) show $R(p_T)$ for single $c$ and $b$ mesons and $R(M)$
and $R(y)$ for $D \overline D$ and $B \overline B$ pairs at RHIC
and LHC.  At RHIC energies, the increase of $R$ with $p_T$ and
$M$ reflects
the integration over the low $x_1$, $x_2$ midrapidity contributions as well as
the growth of $R_{F_2}$ as $x$ approaches the
antishadowing region, reached
at $M \simeq 20$ GeV for $D \overline D$ and $B \overline B$ pairs.  
Note that $R(M)$ is almost identical for $D \overline D$ and
$B \overline B$ production, as it should be since at fixed $M$ the same $x$
values are probed.  The $B$ $p_T$ ratios are 
generally flatter because the change in 
$R_{F_2}(x)$ with $p_T$ is slower than the change in $R_{F_2}(x)$
for $D$ production at the same energy.
The ratio $R(y)$ is nearly constant for $D \overline D$ production
at RHIC, caused by the coincidence of $x_1$ increasing toward the antishadowing
region while $x_2$ decreases into the saturation region.
In contrast, the $B \overline B$ ratio decreases with rapidity at
RHIC since at $y=2$, $x_1$ lies in the EMC region while $x_2 \sim 0.001$.
Figure 17 shows $R(y)$ and $R(M)$ for Drell-Yan pairs at RHIC and LHC.
At RHIC $R(y)$ increases with mass at $y=0$ due to the
increase of $R_{F_2}$ as $x_1=x_2$ grows from 0.01 to 0.03.  At higher
rapidities, $R_{F_2}(x_2)$ moves toward the saturation region while
$R_{F_2}(x_1)$ passes through the EMC region and then rises again as
$x_1 \rightarrow 1$.  Near the edge of
phase space, the transition from the EMC region to the Fermi motion region can
be seen.  The phase space is depleted at higher rapidities for lower
masses.  At the LHC $x_1$ remains in the shadowing region for $y<4$.
The ratios of rapidity integrated mass distributions
rise slowly with mass, reflecting the broad rapidity distributions.

The second parameterization modifies the
valence and sea quark and gluon distributions separately and includes 
evolution with the square of the momentum transfer, $Q^2$, \cite{KJE} 
but is based on an older fit to the data using the
Duke-Owens parton densities \cite{DO}.  
The exact form of $R_{F_2}(x,Q_0)$ is given in Ref.\
\cite{KJE}.  The initial scale is chosen to be $Q_0 = 2$ GeV and the $Q^2$
evolution is done with both the standard Altarelli-Parisi evolution and with
gluon recombination at high density.  The gluon recombination terms do not
strongly alter the evolution.  
In this case, the nuclear parton densities are modified so
that \be q_V^A(x,\mu) & = & R_V(x,\mu) q_V^p(x,\mu) \\  
q_S^A(x,\mu) & = & R_S(x,\mu) q_S^p(x,\mu) \\ g^A(x,\mu) & =
& R_G(x,\mu) g^p(x,\mu) \, \, , \ee where $q_V = u_v + d_v$ is the valence 
quark density and $q_S = 2(\overline u + \overline d + \overline s)$ is
the total
sea quark density and we assume that $R_V$ and $R_S$ affect the individual 
valence and sea
quarks identically.  The ratios were constrained in the model \cite{KJE} by
assuming that $R_{F_2} \approx R_V$ at large $x$ and $R_{F_2} \approx R_S$ at
small $x$ since $xq_V(x,\mu) \rightarrow 0$ as $x \rightarrow 0$.  We use
ansatz 1 for the gluons, $R_{F_2} \approx R_G$ for all $x$ \cite{KJE},
since one might expect more shadowing for the sea quarks, generated from
gluons, at small $x$.
The parton densities satisfy baryon number
conservation $\int_0^1 dx \, q_V^{p,A}(x,\mu) = 3$ and momentum conservation
$\int_0^1 dx \, x(q_V^{p,A}(x,\mu) + q_S^{p,A}(x,\mu) + g^{p,A}(x,\mu)) = 1$
at all $\mu$.  We have used the MRS D$-^\prime$ densities with this
parameterization instead of the original parton densities, leading to some
small deviations in the momentum sum but the general trend is unchanged.

Figure 18(a) shows the ratios $R_V$ (solid curves), $R_S$ (dashed curves), and
$R_G$ (dot-dashed curves) for the minimum, $Q_0$, and maximum, $10$ GeV, 
values of $Q$ for $A=200$.  Outside this range the ratios are fixed
to those at 10 GeV.  The valence quarks show little $Q^2$
evolution, the gluons the greatest.  The sea quarks evolve more slowly than
the gluons and, at $Q=10$ GeV, $R_S \approx R_{F_2}$ of the first
parameterization, eq.\ (1).  The ratios of $AA$ to $pp$ production of
$Q\overline Q$ and Drell-Yan pairs are shown in Fig.\ 18(b)-(d) and Fig.\ 19.
The ratio $R$ for this parameterization is not as straightforward to write down
as in eq.\ (4) except for $gg$ fusion.  For example for $q\overline q
\rightarrow Q \overline Q$, \be
R \propto \frac{R_S(x_1)R_S(x_2)\Sigma_{q=u,d,s}
2q_s(x_1) \overline q_s(x_2) + \left[R_V(x_1)R_S(x_2)
\Sigma_{q=u,d} q_v(x_1) \overline q_s(x_2) + (1\rightarrow 
2)\right]}{\Sigma_{q=u,d,s} 2q_s(x_1) \overline q_s(x_2)
+ \left[\Sigma_{q=u,d} q_v(x_1) \overline q_s(x_2) +
(1\rightarrow 2)\right]} \, \, , \ee where $q_s$ is a generic sea quark
distribution and the $\mu$ dependence has been suppressed.  
The fast evolution of the gluons has the strongest effect on the charm and
bottom production since $Q \overline Q$ production by gluons is dominant.  
In fact, for $b
\overline b$ production at RHIC energies, the shadowing effect has nearly
vanished in $R(p_T)$.  The ratios for Au+Au to $pp$ are shown in Fig.\
18(b)-(d) as a function of quark $p_T$ and pair mass and rapidity.  The trends
are the same as for the first parameterization but, overall, the $Q \overline
Q$ distributions are not as strongly modified since the effect decreases
for increasing $p_T$ and $M$.  As seen in a comparison of Fig.\ 19 with
Fig.\ 17, the shadowing
effect is actually stronger for low mass Drell-Yan production with this 
parameterization due to the strong sea quark shadowing at $Q_0$.

In either case, the reduction in $AA$ yield relative to $pp$ due to shadowing
is generally not larger than a factor of two,
depending on the shadowing model.  It is important to note that the total 
depletion is a dependent on both $x$ and $Q$ and is not a constant factor
as a function of $p_T$, $M$ and $y$ in either model.

\clearpage
\begin{table}
\begin{center}
\begin{tabular}{|c|c|c|c|} \hline
& \multicolumn{3}{|c|}{$N_{ll}$} \\ \hline
Source  & $M_{ll}=2$ GeV & $M_{ll}=4$ GeV & $M_{ll}=6$ GeV \\ \hline
$D \overline D_{\rm uncorr}$ & 1.94$\times 10^{-1}$ & 4.5$\times 10^{-2}$ &
1.67$\times 10^{-2}$ \\ \hline
$D \overline D_{\rm corr}$ & 2.31$\times 10^{-2}$ & 2.10$\times 10^{-3}$ &
3.10$\times 10^{-4}$ \\ \hline
$B \overline B$ & 1.38$\times 10^{-4}$ & 6.85$\times 10^{-5}$ &
2.52$\times 10^{-5}$ \\ \hline
DY & 5.29$\times 10^{-4}$ & 3.86$\times 10^{-5}$ &
8.06$\times 10^{-6}$ \\ \hline
$l^+ l^-_{\rm th}$ & 9.07$\times 10^{-4}$ & 7.6$\times 10^{-6}$ &
1.4$\times 10^{-7}$ \\ \hline
$D \overline D_{\rm th}$ & 9.29$\times 10^{-4}$ &
5.18$\times 10^{-6}$ & $-$ \\ \hline
\end{tabular}
\end{center}
\caption[]{ Number of lepton pairs per event
from each of our sources in central Au+Au collisions at RHIC.}
\end{table}
\clearpage

\begin{table}
\begin{center}
\begin{tabular}{|c|c|c|c|} \hline
& \multicolumn{3}{|c|}{$N_{ll}$} \\ \hline
Source  & $M_{ll}=2$ GeV & $M_{ll}=4$ GeV & $M_{ll}=6$ GeV \\ \hline
$D \overline D_{\rm uncorr}$ & 4.80$\times 10^{2}$ & 1.90$\times 10^{2}$ &
1.07$\times 10^{2}$ \\ \hline
$D \overline D_{\rm corr}$ & 1.40$\times 10^{0}$ & 1.69$\times 10^{-1}$ &
2.95$\times 10^{-2}$ \\ \hline
$B \overline B$ & 2.05$\times 10^{-2}$ & 1.15$\times 10^{-2}$ &
4.98$\times 10^{-3}$ \\ \hline
DY & 6.90$\times 10^{-3}$ & 7.83$\times 10^{-4}$ &
2.06$\times 10^{-4}$ \\ \hline
$l^+ l^-_{\rm th}$ & 1.43$\times 10^{-2}$ & 4.68$\times 10^{-4}$ &
3.26$\times 10^{-5}$ \\ \hline
$D \overline D_{\rm th, corr}$ & 4.8$\times 10^{-2}$ &
1.02$\times 10^{-3}$ & 3.57$\times 10^{-5}$ \\ \hline
$D \overline D_{\rm th, uncorr}$ & 1.42$\times 10^{0}$ &
4.32$\times 10^{-1}$ & 2.03$\times 10^{-1}$ \\ \hline
\end{tabular}
\end{center}
\caption[]{ Number of lepton pairs per event from each of our sources in
central Pb+Pb collisions at LHC.}
\end{table}
\clearpage

\begin{table}
\begin{center}
\begin{tabular}{|c|c|c|c|c|} \hline
$\sqrt{s}$  & \multicolumn{2}{|c|}{$ 200$ GeV} & \multicolumn{2}{|c|}{$ 5.5$
TeV} \\ \hline
Source & $\langle M_{ll} \rangle$ (GeV) & $\langle p_{T, ll} \rangle$ (GeV) &
$\langle M_{ll} \rangle$ (GeV) & $\langle p_{T, ll} \rangle$ (GeV) \\ \hline
$D \overline D_{\rm uncorr}$ & 3.73 & 0.84 & 4.57 & 1.00 \\ \hline
$D \overline D_{\rm corr}$ & 2.41 & 0.77 & 3.03 & 0.84 \\ \hline
$B \overline B$ & 3.95 & 1.91 & 4.20 & 2.05  \\ \hline
DY & 2.45 & 1.68 & 2.68 & 2.45 \\ \hline
$l^+ l^-_{\rm th}$ & 2.44 & 1.06 & 2.61 & 1.26 \\ \hline
$D \overline D_{\rm th, corr}$ & 2.39 & 0.85 & 2.52 & 0.98 \\ \hline
$D \overline D_{\rm th, uncorr}$ & - & - & 2.28 & 0.97 \\ \hline
\end{tabular}
\end{center}
\caption[]{ Average lepton pair mass (for $M_{ll} > 2$ GeV)
and transverse momentum at RHIC and LHC.  For
the Drell-Yan and thermal lepton pairs, the average $p_T$'s are calculated
for $M_{ll}> 2$ GeV only while all the average $p_T$'s from heavy quark decays
are for all masses.}
\end{table}
\clearpage

\begin{table}
\begin{center}
\begin{tabular}{|c|c|c|c|c|c|c|} \hline
 & \multicolumn{2}{|c|}{$e^+e^-$} & \multicolumn{2}{|c|}{$\mu^+\mu^-$} &
\multicolumn{2}{|c|}{$e^\pm \mu^\mp$} \\ \hline
Source  & \% Acc. & $\langle M_{ee} \rangle$ (GeV) & \% Acc. &
$\langle M_{\mu\mu} \rangle$ (GeV) & \% Acc. & $\langle M_{e\mu} \rangle$ 
(GeV) \\ \hline
$D \overline D_{\rm uncorr}$ & 0.01 & 2.75 & 0.07 & 2.53 & 0.084 & 3.2 
\\ \hline
$D \overline D_{\rm corr}$ & 0.032 & 3.05 & 0.42 & 2.84 & 0.08 & 3.74 \\ \hline
$B \overline B$ & 0.60 & 4.53 & 1.7 & 3.56 & - & -  \\ \hline
DY & 0.16 & 3.15 & 3.7 & 2.64  & - & - \\ \hline
$l^+ l^-_{\rm th}$ & 0.23 & 2.51 & 3.3 & 2.18 & - & - \\ \hline
$D \overline D_{\rm th}$ & 0.014 & 2.47 & 0.12 & 2.58 & - & - \\ \hline
\end{tabular}
\end{center}
\caption[]{Percentage of lepton pairs with $M_{ll} > 2$ GeV
accepted in the PHENIX detector and their average pair mass.   
We have included the central electron arms, the
forward muon arm, and a combination of electrons and muons from both
detectors.}
\end{table}
\clearpage

\begin{table}
\begin{center}
\begin{tabular}{|c|c|c|c|c|} \hline
  & \multicolumn{2}{|c|}{$e^+e^-$} & \multicolumn{2}{|c|}{$\mu^+\mu^-$}
\\ \hline
  & \% Acc. & $\langle M_{ee} \rangle$ (GeV) & \% Acc. &
$\langle M_{\mu\mu} \rangle$ (GeV) \\ \hline
$D \overline D_{\rm uncorr}$ & 0.093 & 2.69 & 0.12 & 2.58 \\ \hline
$D \overline D_{\rm corr}$ & 0.44 & 3.10 & 0.67 & 2.92 \\ \hline
$B \overline B$ & 4.42 & 4.33 & 3.58 & 4.01 \\ \hline
DY & 3.6 & 2.62 & 5.27 & 2.58  \\ \hline
$l^+ l^-_{\rm th}$ & 4.9 & 2.55 & 4.76 & 2.22 \\ \hline
$D \overline D_{\rm th, corr}$ & 0.46 & 2.52 & 0.29 & 2.34 \\ \hline
$D \overline D_{\rm th, uncorr}$ & 0.33 & 2.70 & 0.056 & 2.49 \\ \hline
\end{tabular}
\end{center}
\caption[]{Percentage of lepton pairs with $M_{ll} > 2$ GeV
accepted in the ALICE detector and their average pair mass. 
We have included the central detector and the proposed forward muon arm.}
\end{table}
\clearpage

\begin{center}
{\bf Figure Captions}
\end{center}
\vspace{0.25in}

\noi Figure 1.  The lepton pair mass distributions are given in (a) for
central Au+Au collisions at RHIC and (b) for central Pb+Pb collisions at LHC.
The contributions are:  Drell-Yan (dashed) and
thermal dilepton (dot-dashed-dashed) production and thermal $D \overline D$
decays (dotted), as well as initial correlated (dot-dashed) $D \overline D$ and
$B \overline B$ (dot-dot-dashed) production and decay.  Note that lepton pairs
from uncorrelated initial $D \overline D$ decays (as well as uncorrelated
thermal $D \overline D$ decays at the LHC) have not been included here 
but are a very large contribution to the continuum.

\vspace{0.2in}

\noi Figure 2. The NLO Drell-Yan pair production rate in central
Au+Au collisions at $\sqrt{s} = 200$ GeV.
The pair $p_T$ (a), mass (b), and rapidity (c) distributions are shown.
The $p_T$ distributions are given at $y=0$ (solid) and
$y=2$ (dashed) for $M=2$, 4, and 6 GeV (the upper, middle, and lower
sets of curves respectively).  In (c) the rapidity distributions are calculated
for $M=2$ GeV (solid), 4 GeV (dashed), and 6 GeV (dot-dashed).

\vspace{0.2in}

\noi Figure 3. The same as Fig.\ 2 for central Pb+Pb collisions at
$\sqrt{s} = 5.5$ TeV.

\vspace{0.2in}

\noi Figure 4.  The rate of initial $c \overline c$ (solid) and $D
\overline D$ (dashed) pair production to NLO in central Au+Au collisions at 
$\sqrt{s} = 200$ GeV.
The pair $p_T$ (a), mass (b), rapidity gap (c), and pair rapidity (d)
distributions are shown.   Additionally, the lepton pair $p_T$, mass, and
rapidity distributions from correlated (dot-dashed) and uncorrelated (dotted)
$D \overline D$ pair decays are shown for central Au+Au collisions.

\vspace{0.2in}

\noi Figure 5.  The same as Fig.\ 4 for central Pb+Pb collisions at
$\sqrt{s} = 5.5$ TeV.

\vspace{0.2in}

\noi Figure 6.  The same as Fig.\ 4 for the $b \overline b$ and $B \overline B$
production rate to NLO in central Au+Au collisions at $\sqrt{s} = 200$ GeV.

\vspace{0.2in}

\noi Figure 7.  The same as Fig.\ 6 for  central Pb+Pb collisions at
$\sqrt{s} = 5.5$ TeV.

\vspace{0.2in}

\noi Figure 8.  The thermal lepton pair rate in central Au+Au collisions
for $M \geq 2$ GeV and $\sqrt{s} = 200$ GeV.   The lepton pair $p_T$ (a),
mass (b), rapidity gap (c), and rapidity (d) distributions are shown.
The $p_T$, rapidity gap, and rapidity distributions are given for $M=2$
(solid), 4 (dashed), and 6 (dot-dashed) GeV.

\vspace{0.2in}

\noi Figure 9.  The same as Fig.\ 8 for central Pb+Pb collisions at
$\sqrt{s} = 5.5$ TeV.

\vspace{0.2in}

\noi Figure 10.  The thermal $c \overline c$ (solid) and $D
\overline D$ (dashed) pair production rate in central Au+Au collisions
at $\sqrt{s} = 200$ GeV.
The pair $p_T$ (a), mass (b), rapidity gap (c), and rapidity (d)
distributions are given.  The lepton pairs from thermal $D \overline D$
decays are shown in the dot-dashed curves.

\vspace{0.2in}

\noi Figure 11.  The same as Fig.\ 10 for central Pb+Pb collisions at
$\sqrt{s} = 5.5$ TeV.  Uncorrelated thermal $D \overline D$ decays are also
included (dotted curves).

\vspace{0.2in}

\noi Figure 12.  The contributions to the dilepton spectrum in central
Au+Au collisions at $\sqrt{s} = 200$ GeV for pairs with
$M =2$, 4, and 6 GeV.  The $p_T$ distributions are given in (a), (c), and
(e) while the pair rapidity distributions are shown in (b), (d), and (f).
The distributions from Drell-Yan (dashed) and
thermal dilepton  (solid) production and thermal $D \overline D$ decays
(dotted), the initial correlated
(dot-dashed) and uncorrelated (dot-dashed-dashed) $D \overline D$ and
$B \overline B$ (dot-dot-dashed) production and decay are included.

\vspace{0.2in}

\noi Figure 13.  The same as Fig.\ 12 for central Pb+Pb collisions at
$\sqrt{s} = 5.5$ TeV.  In addition to the contributions shown in Fig.\ 12, 
uncorrelated
thermal $D \overline D$ decays are shown in the dot-dot-dashed-dashed curves.

\vspace{0.2in}

\noi Figure 14.  The mass distributions of pairs
accepted into the PHENIX detector in central Au+Au collisions at RHIC.  The
acceptance cuts are shown for pairs in
(a) the central electron detector, (b) the
forward muon arm, and (c) pairs formed when an electron is
accepted into one of the PHENIX
central arms and an opposite sign muon is accepted into the forward arm.
The contributions are:   Drell-Yan (dashed) and
thermal dilepton (solid) production and thermal $D \overline D$
(dotted), initial correlated
(dot-dashed) and uncorrelated (dot-dashed-dashed) $D \overline D$, and initial
$B \overline B$ (dot-dot-dashed) production and decay.  In (c), only the mass
distributions for correlated and uncorrelated $D \overline D \rightarrow
e^\pm \mu^\mp X$ decays are shown.

\vspace{0.2in}

\noi Figure 15.  The same as Fig.\ 14 for pairs accepted into the ALICE
detector in central Pb+Pb collisions at LHC.  The accepted distributions 
are given for (a) the central detector and (b) the proposed forward muon arm.
In addition to the contributions shown in Fig.\ 14, uncorrelated
thermal $D \overline D$ decays are shown in the dot-dot-dashed-dashed curves.

\vspace{0.2in}

\noi Figure 16.  (a) The shadowing function $R_{F_2}(x)$ for $A=197$.  We also
show the charm and bottom Au+Au to $pp$ production ratios as a function of (b)
heavy quark $p_T$, (c) $Q \overline Q$ invariant mass and (d) $Q \overline Q$
pair rapidity.
Charm production is given in the solid curve at RHIC and the dashed curve
at LHC.  Bottom production is shown in the dot-dashed curve for RHIC and the
dotted curve for LHC.  Both Au+Au and $pp$ production is calculated at
$\sqrt{s} = 200$ GeV for RHIC and $\sqrt{s} = 5.5$ TeV at the LHC.

\vspace{0.2in}

\noi Figure 17.  The ratio of Drell-Yan rapidity and mass distributions in
Au+Au to $pp$ collisions at $\sqrt{s} = 200$ GeV at RHIC and 5.5 TeV at the
LHC.  The ratio $R(y)$ is given for $M$=2 (solid), 4 (dashed) and 6 
(dot-dashed) GeV for RHIC (a) and LHC (c).  The mass $R(M)$ is also given 
for RHIC (b) and LHC (d).

\vspace{0.2in}

\noi Figure 18.  (a) The shadowing functions $R_V(x)$ (solid), $R_S(x)$
(dashed) and $R_g(x)$ (dot-dashed) for $Q=2$ (lower curves) and 10 GeV (upper
curves) for $A=200$.  We also
show the charm and bottom Au+Au to $pp$ production ratios as in Fig.\
16 but with the shadowing functions given in (a).

\vspace{0.2in}

\noi Figure 19.  The same as in Fig.\ 17 with the shadowing functions as in
Fig.\ 18(a).

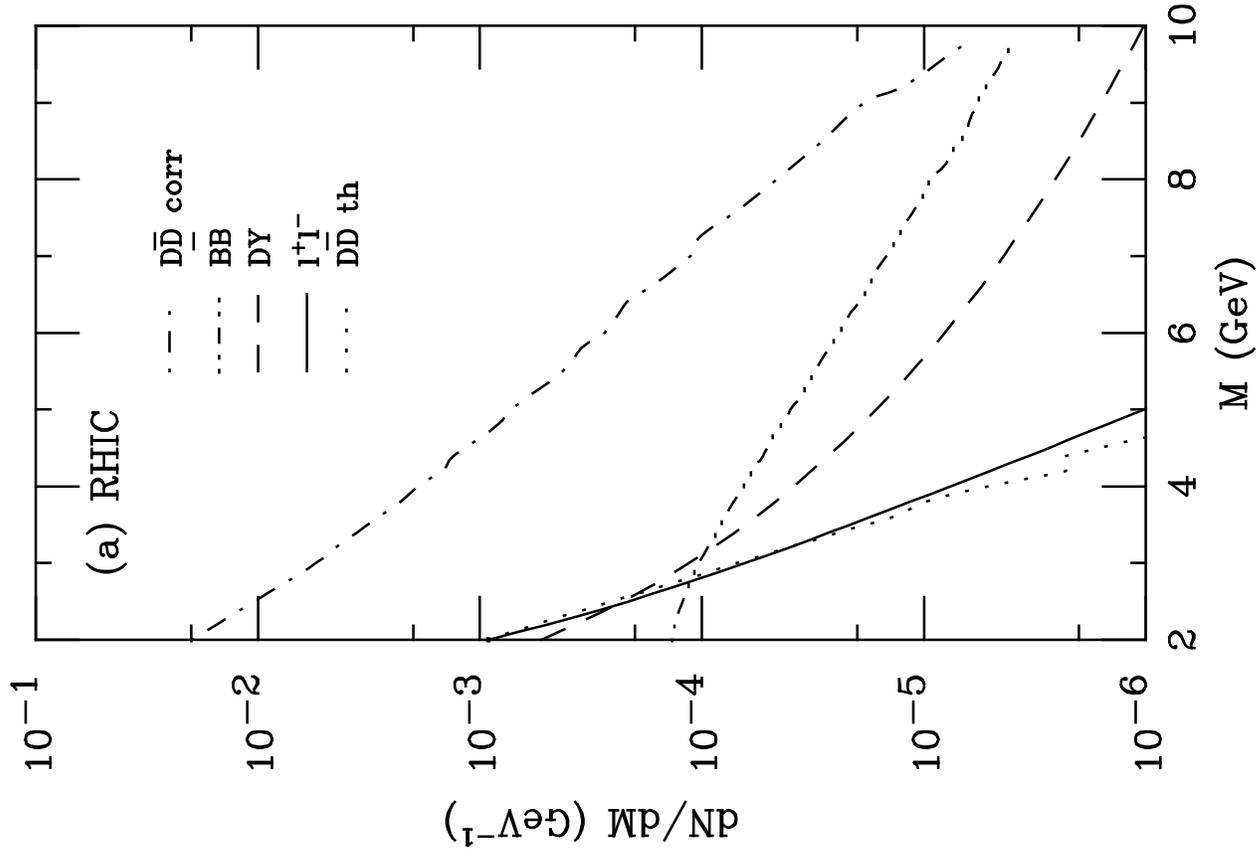
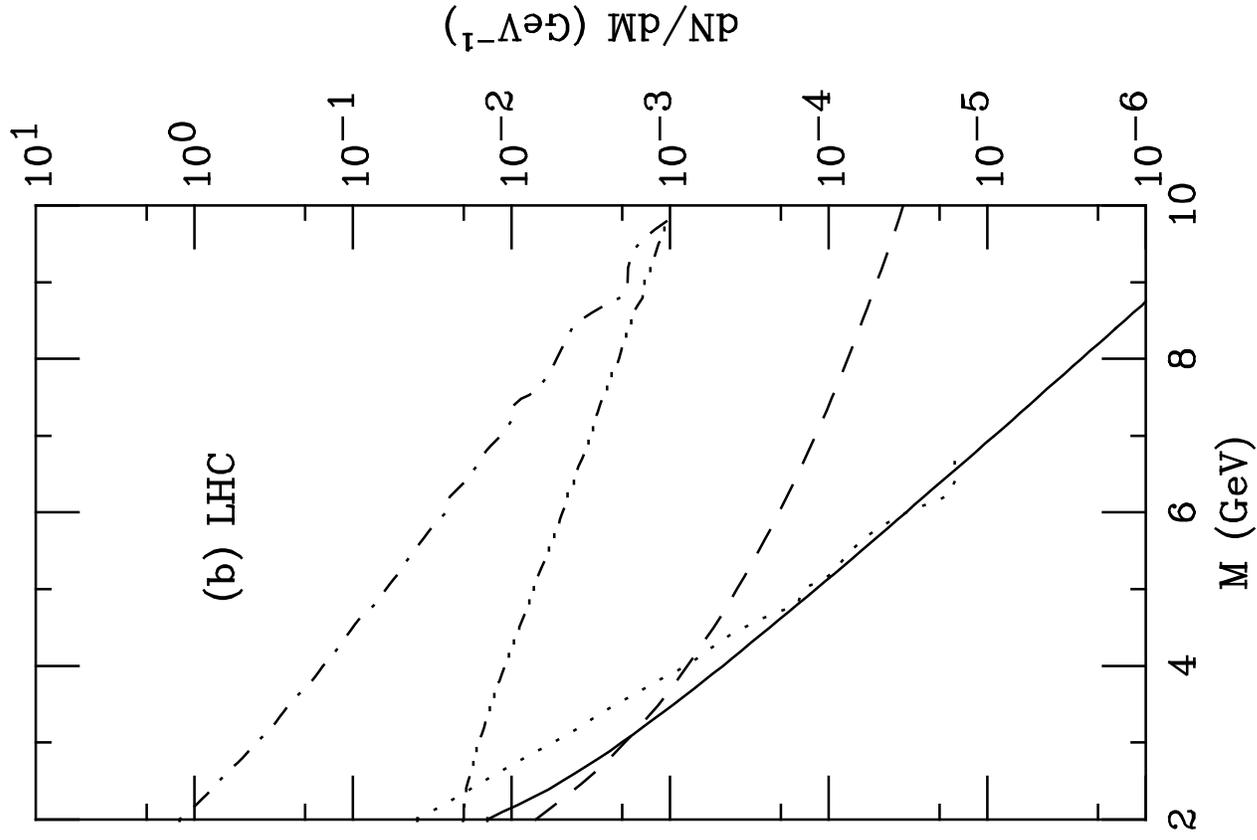
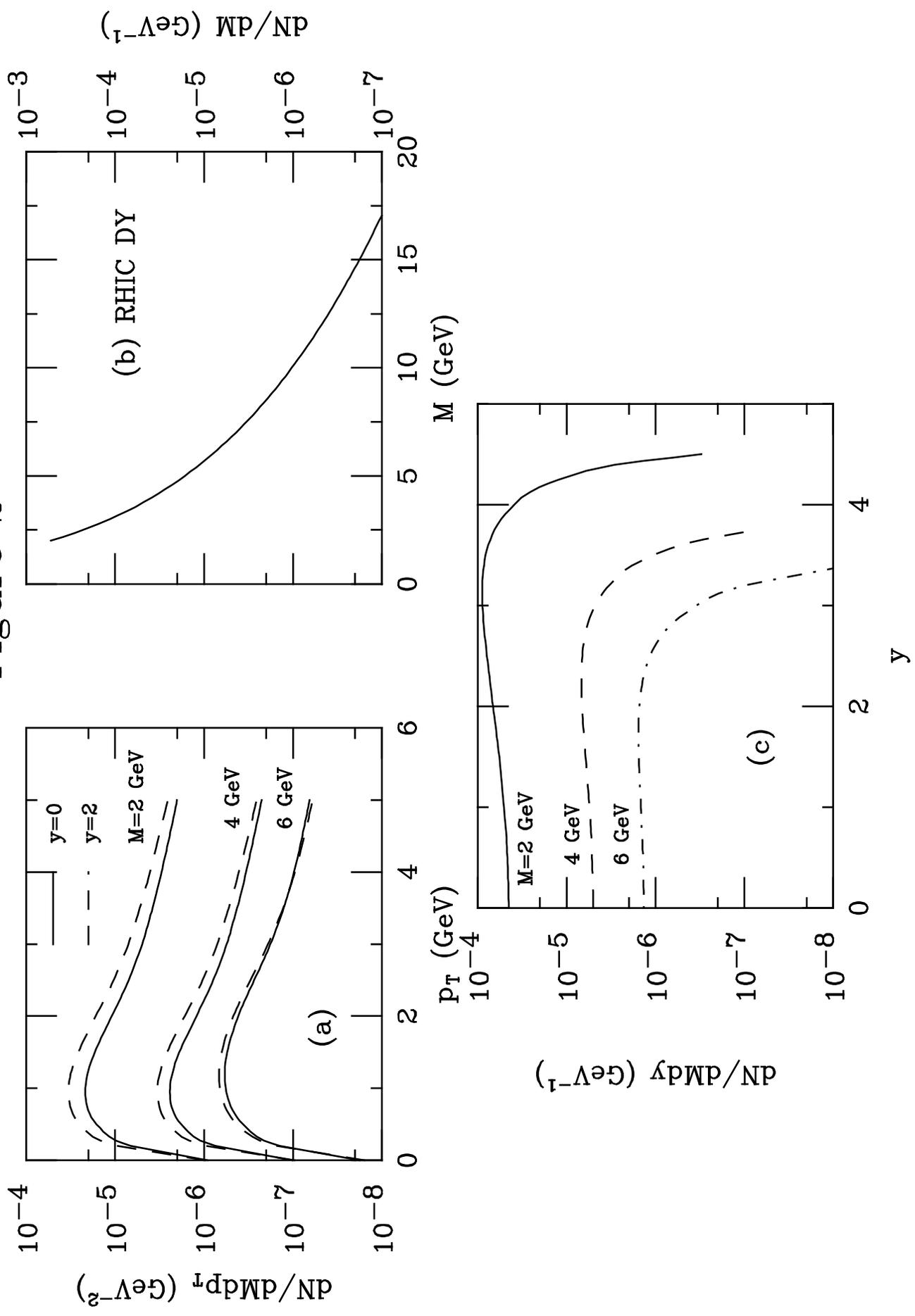
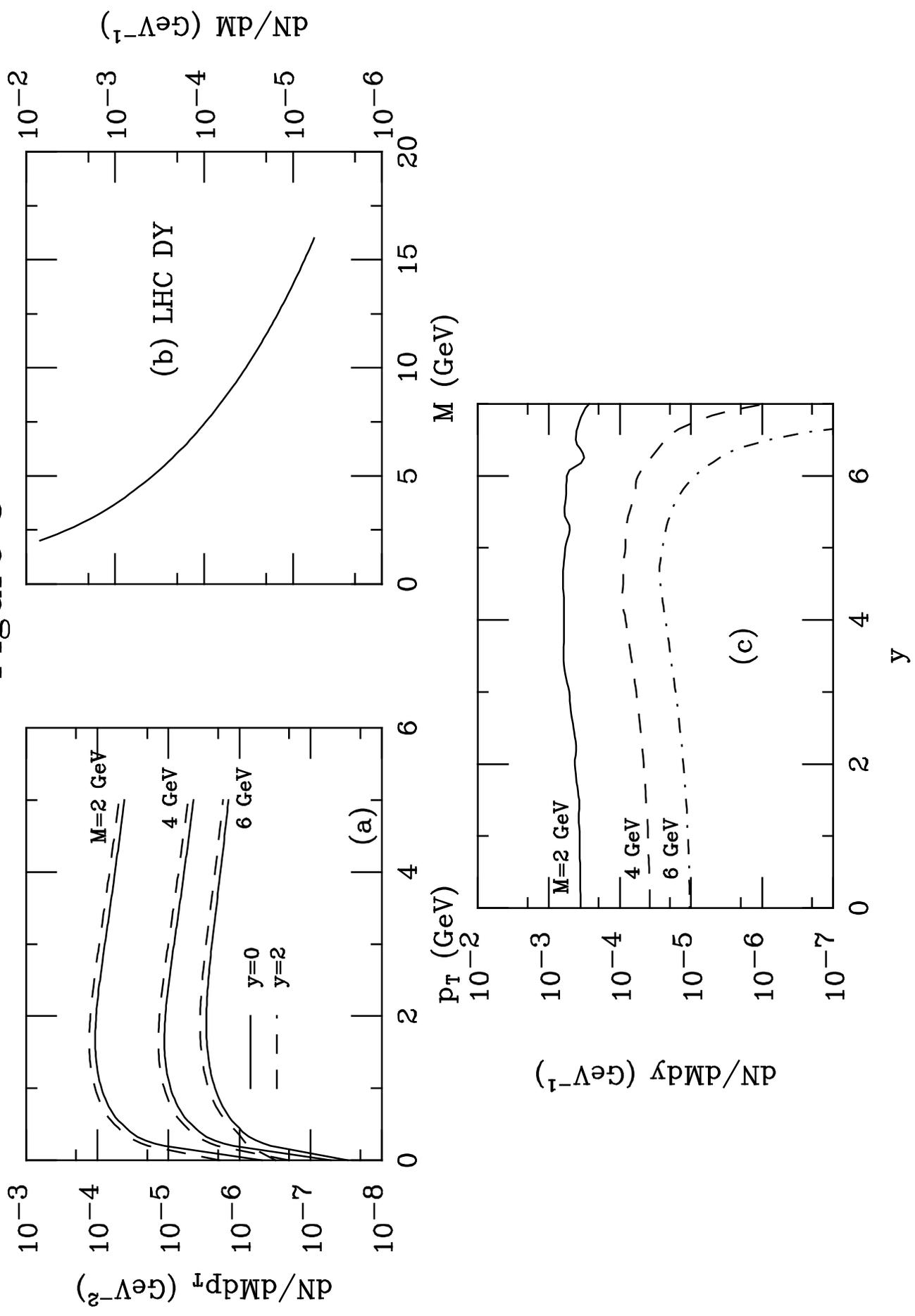
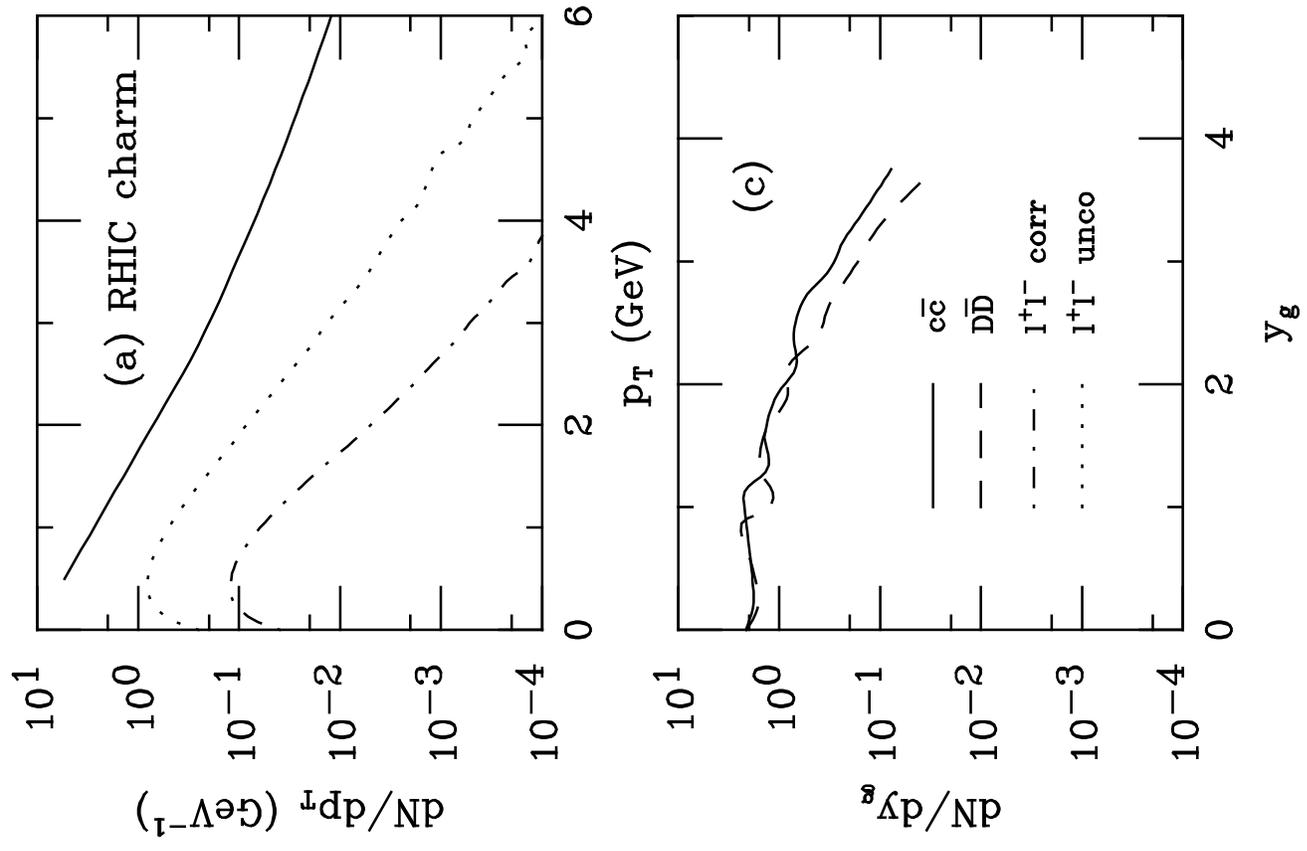
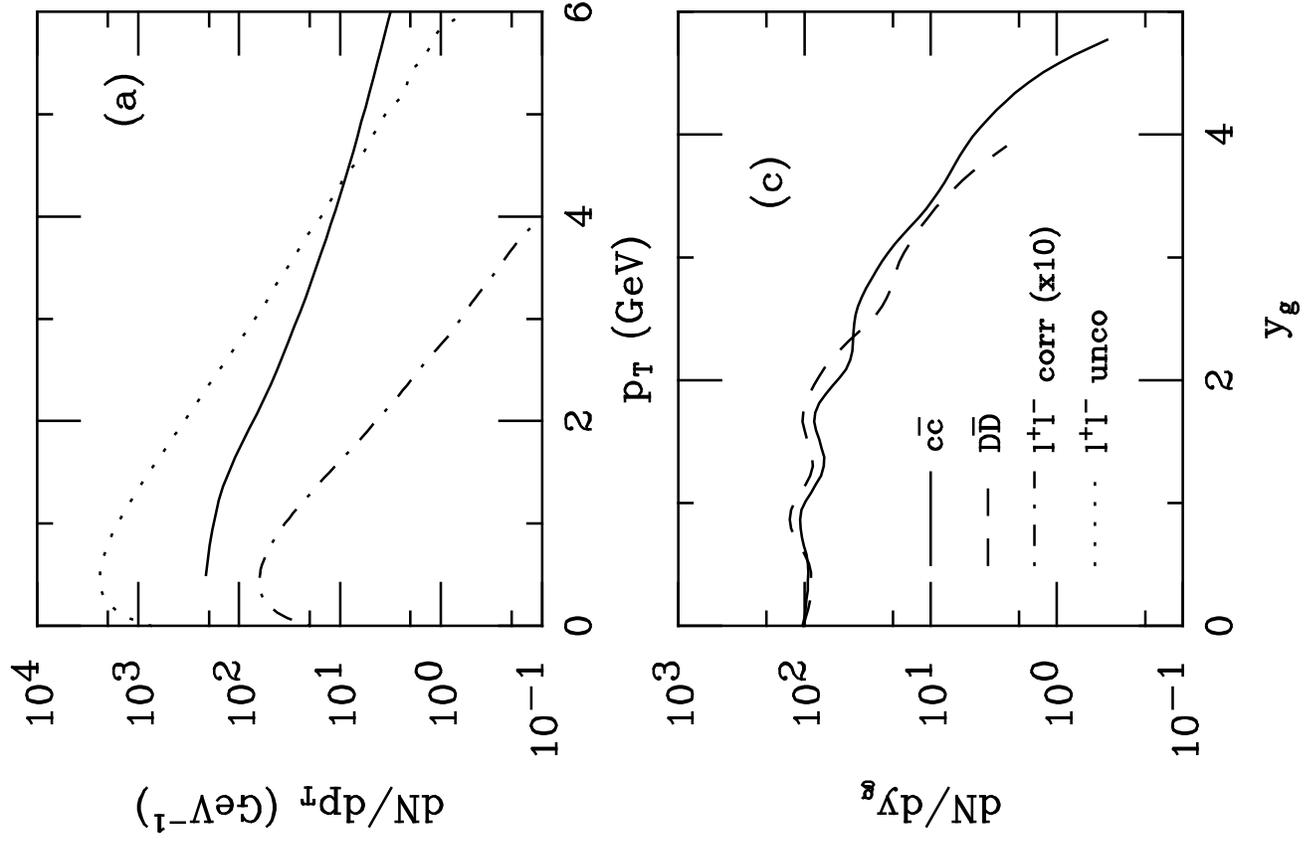
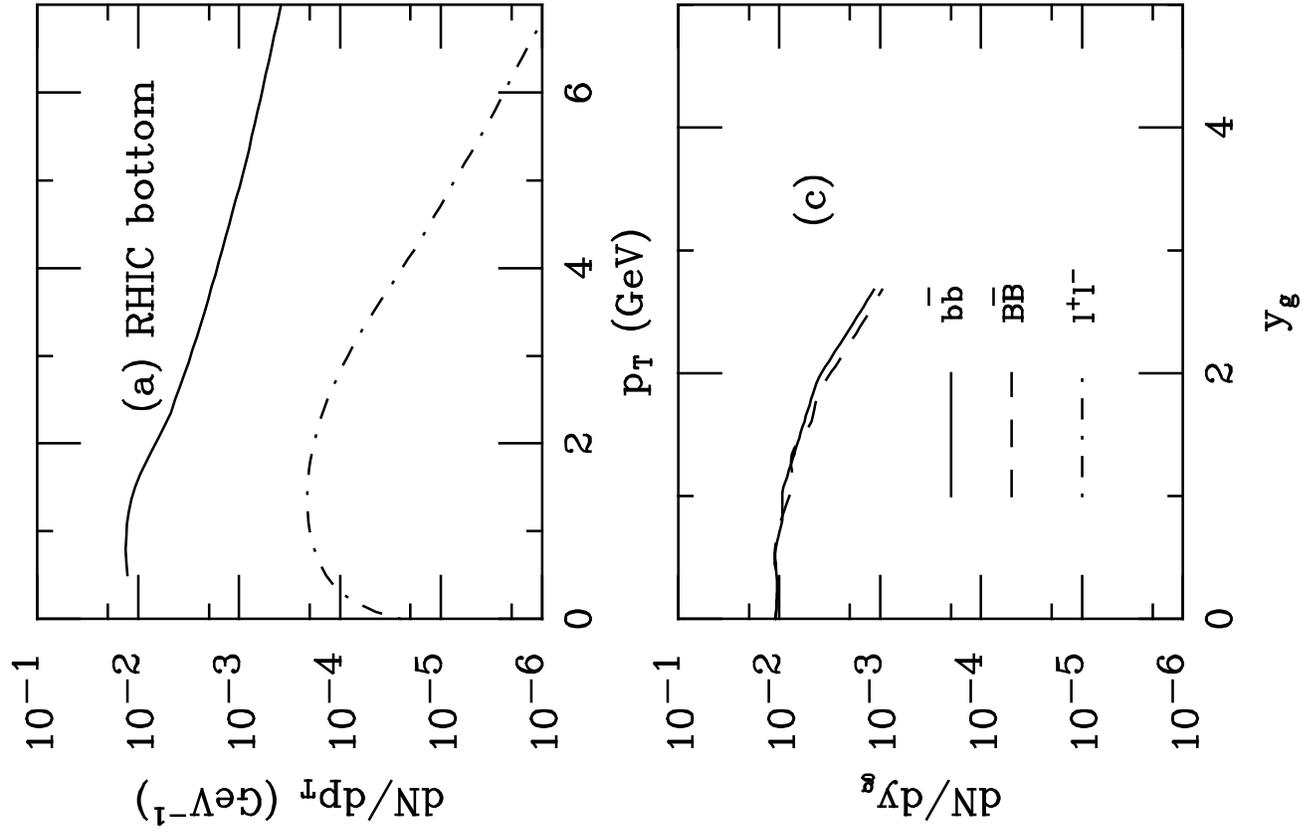
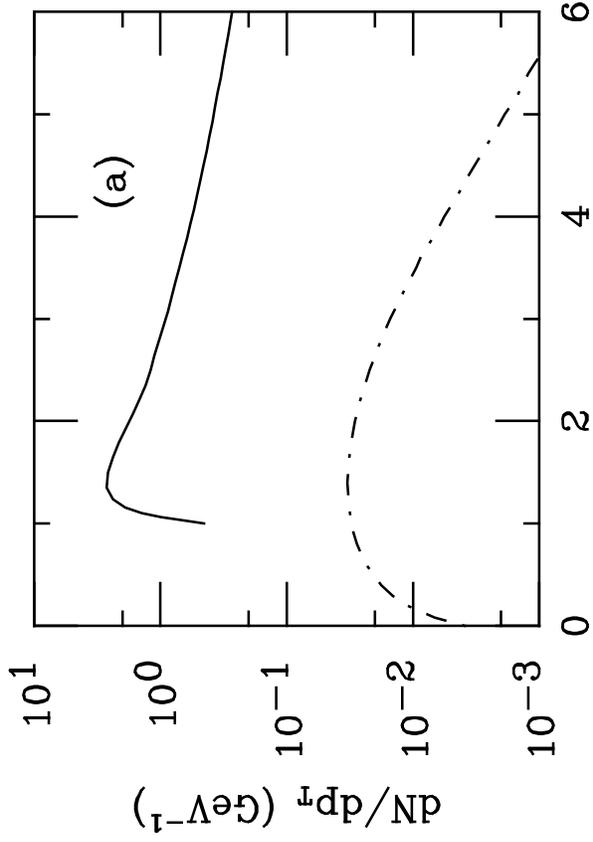
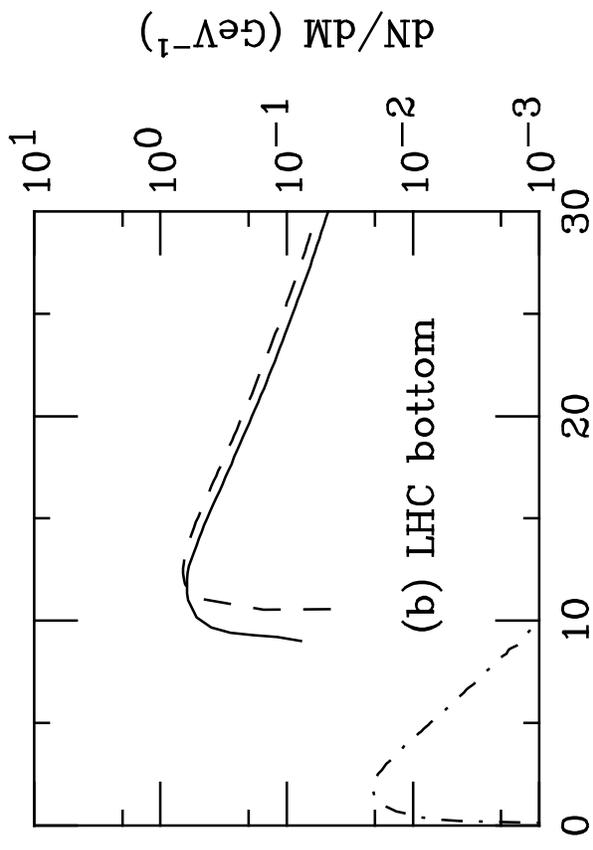
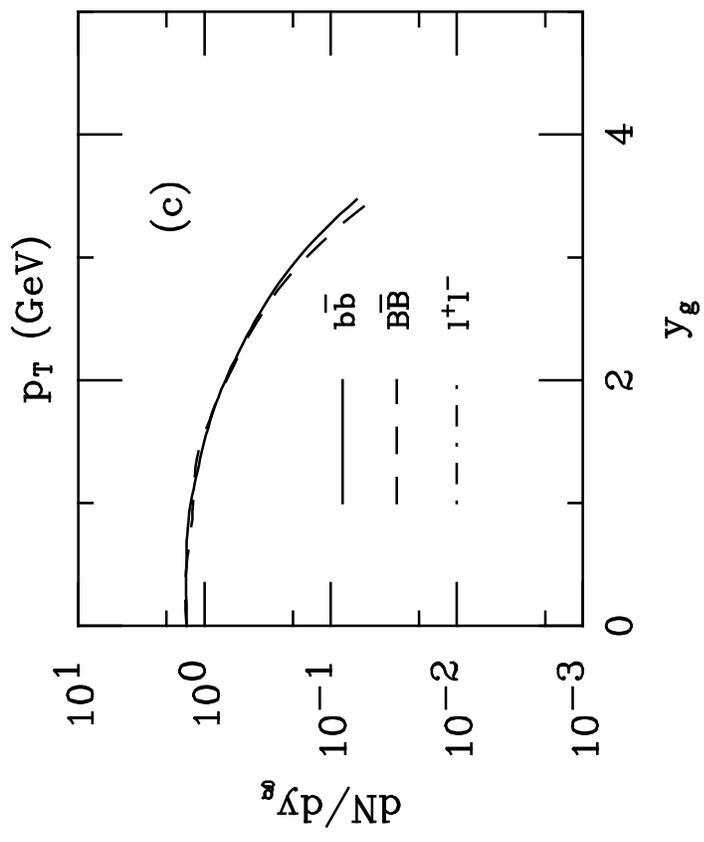
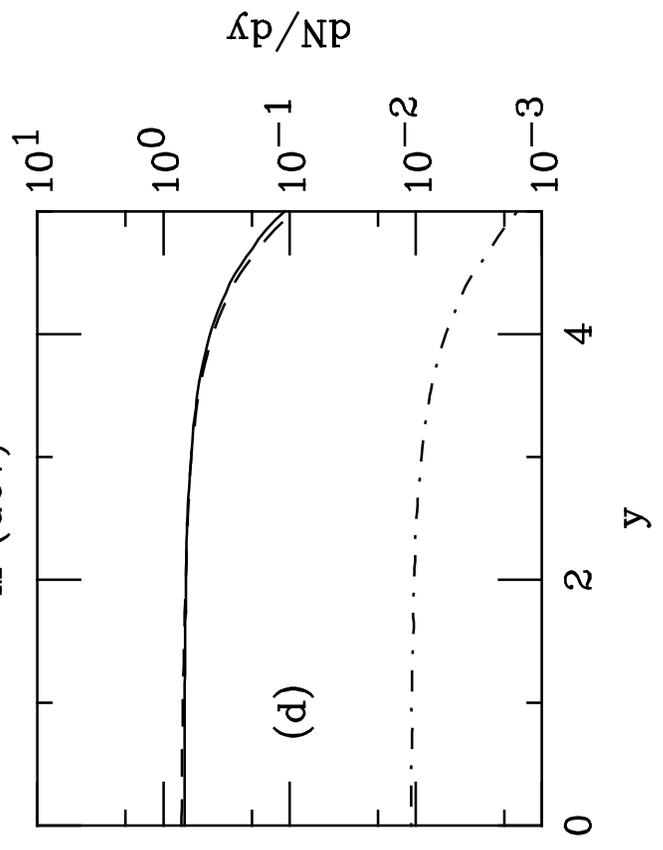
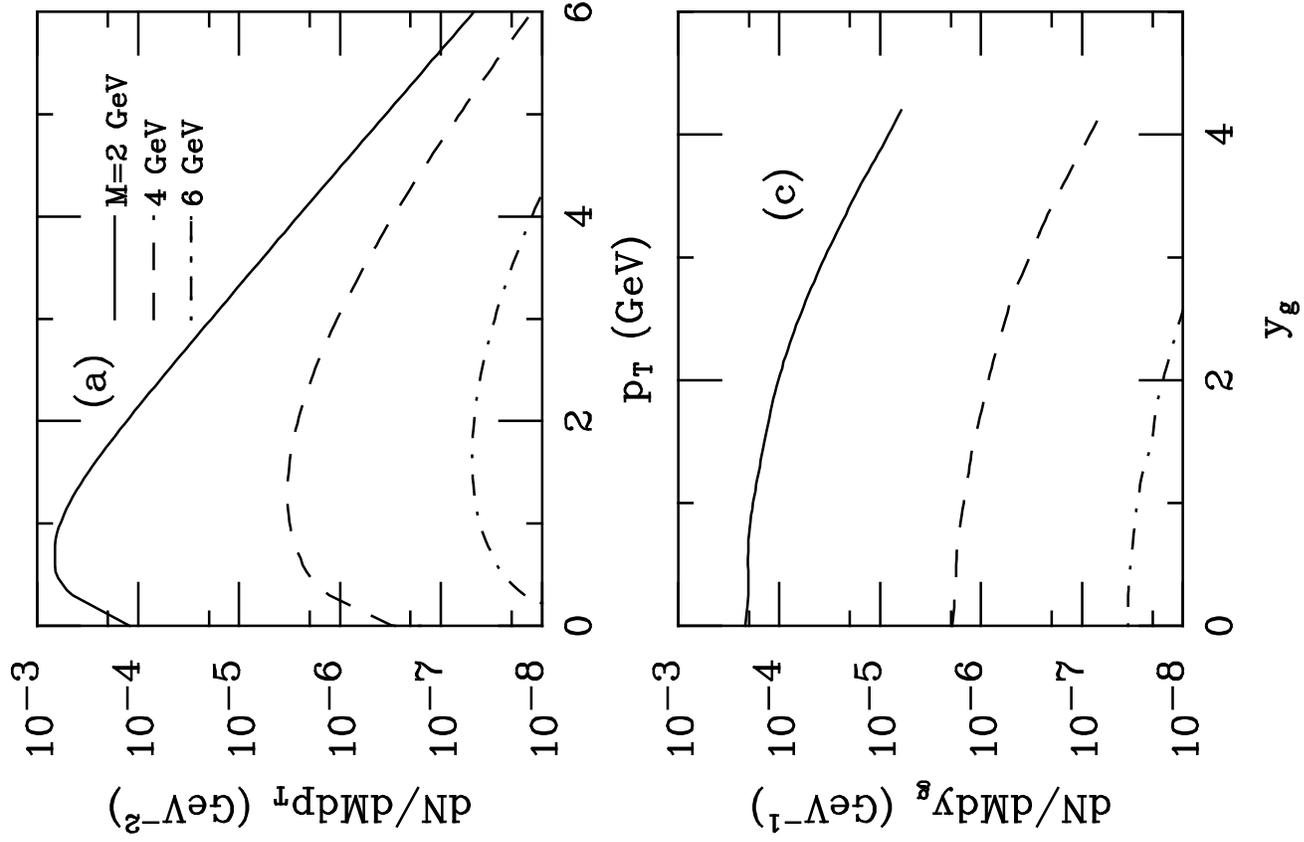
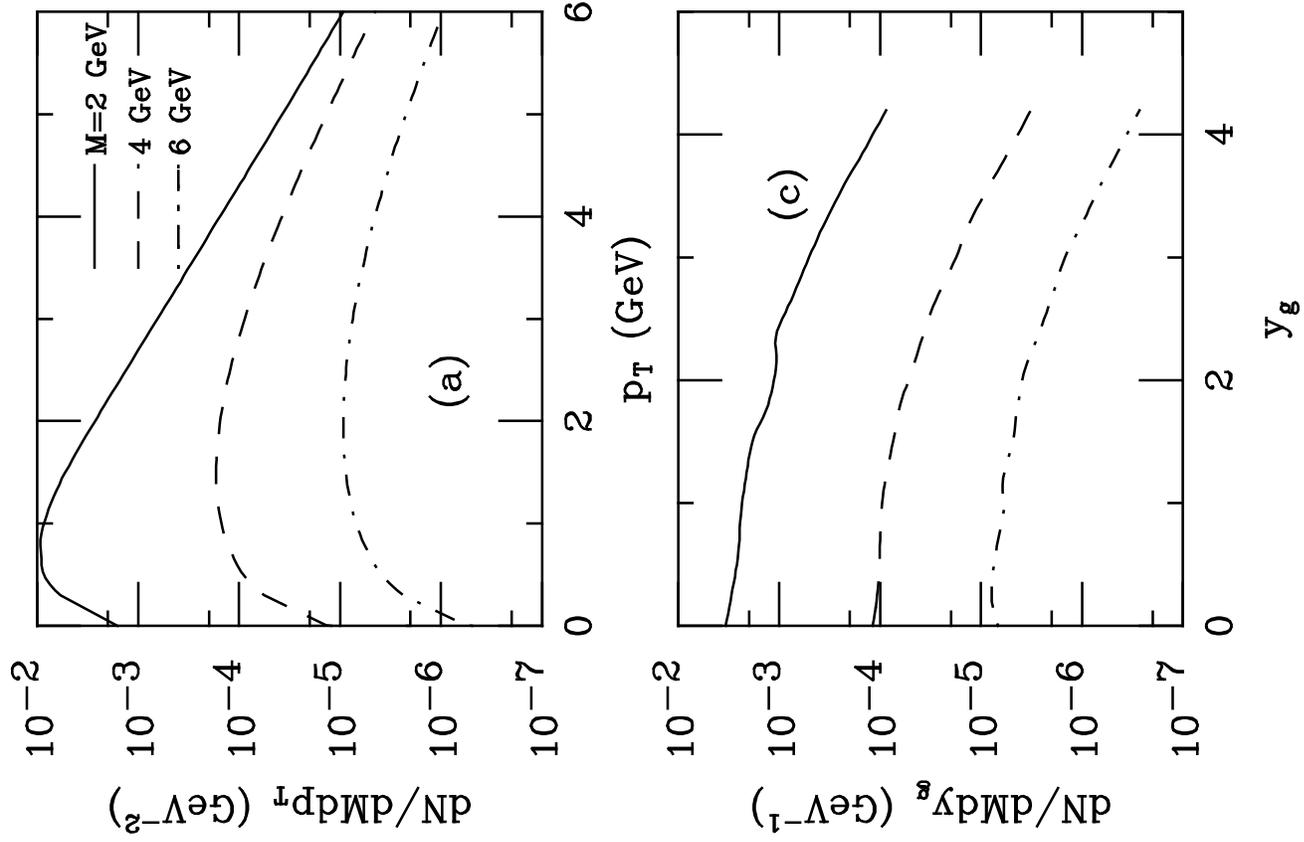
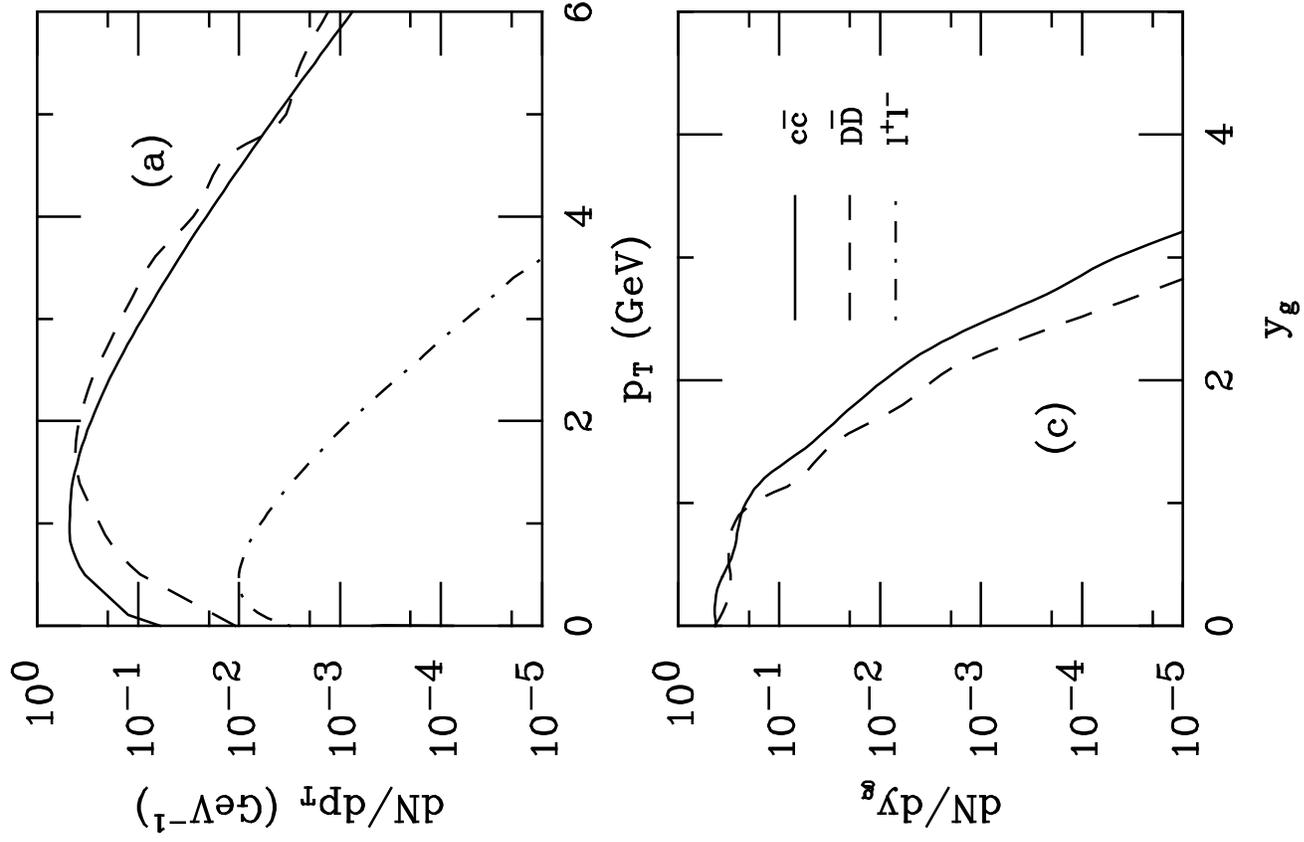
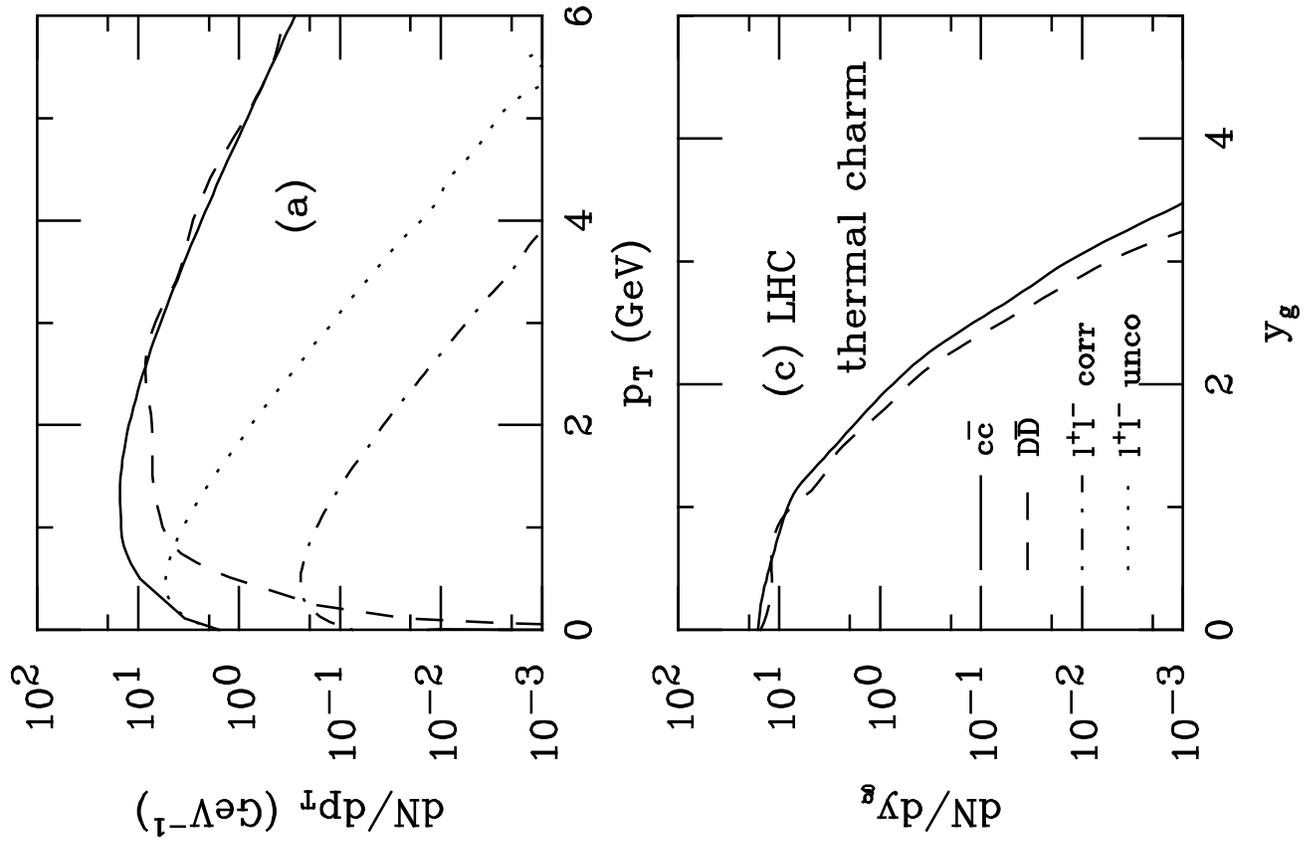
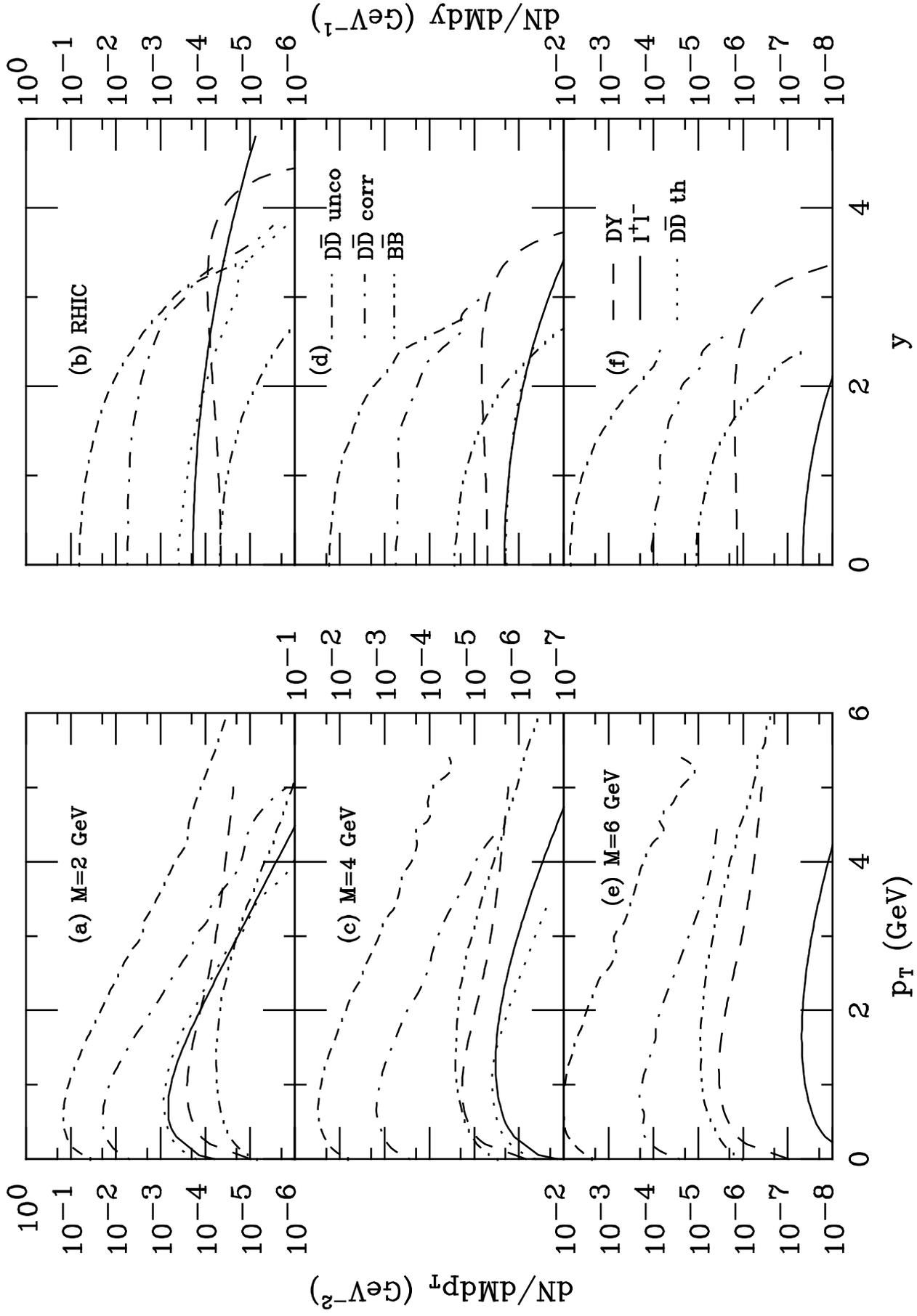
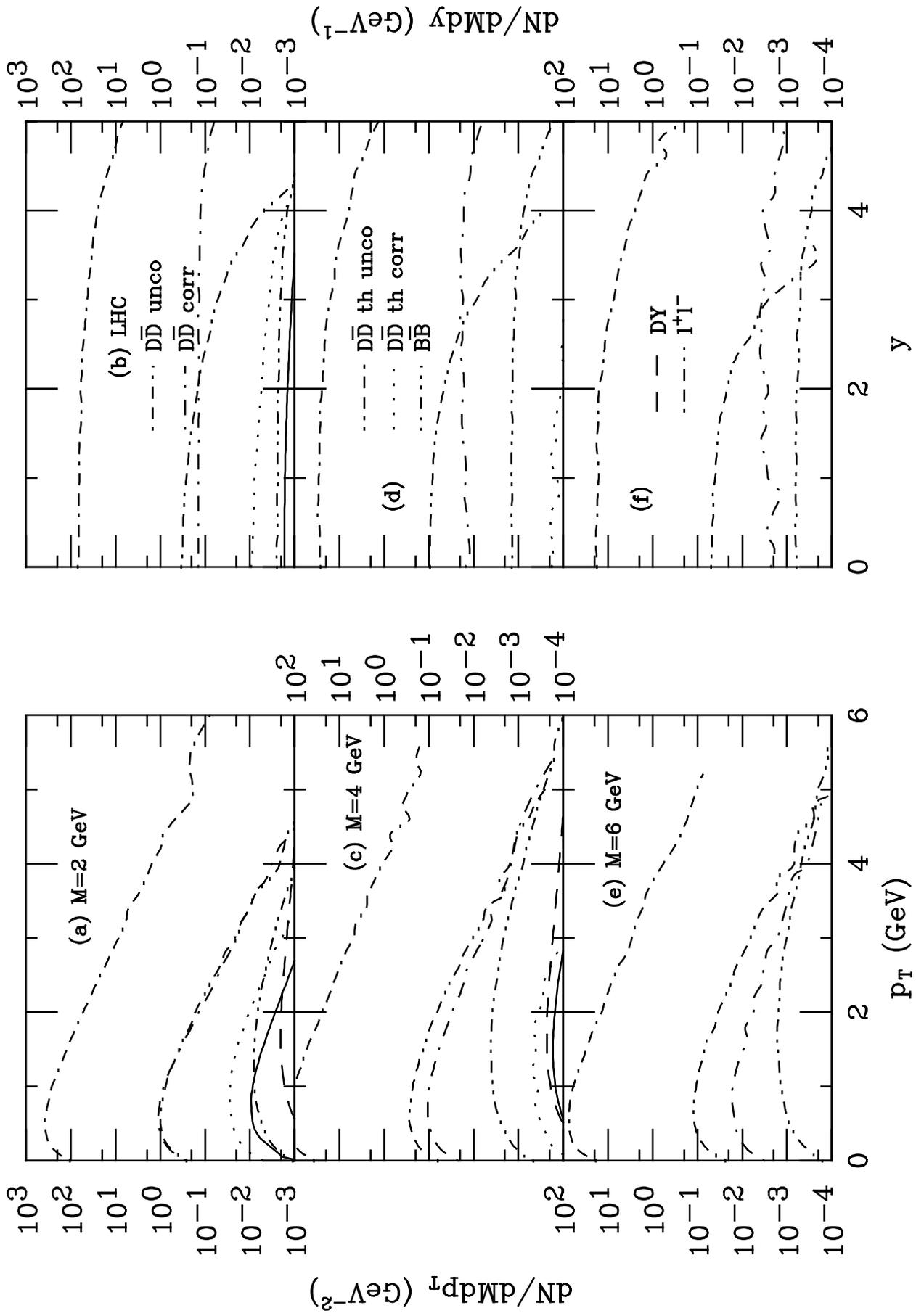
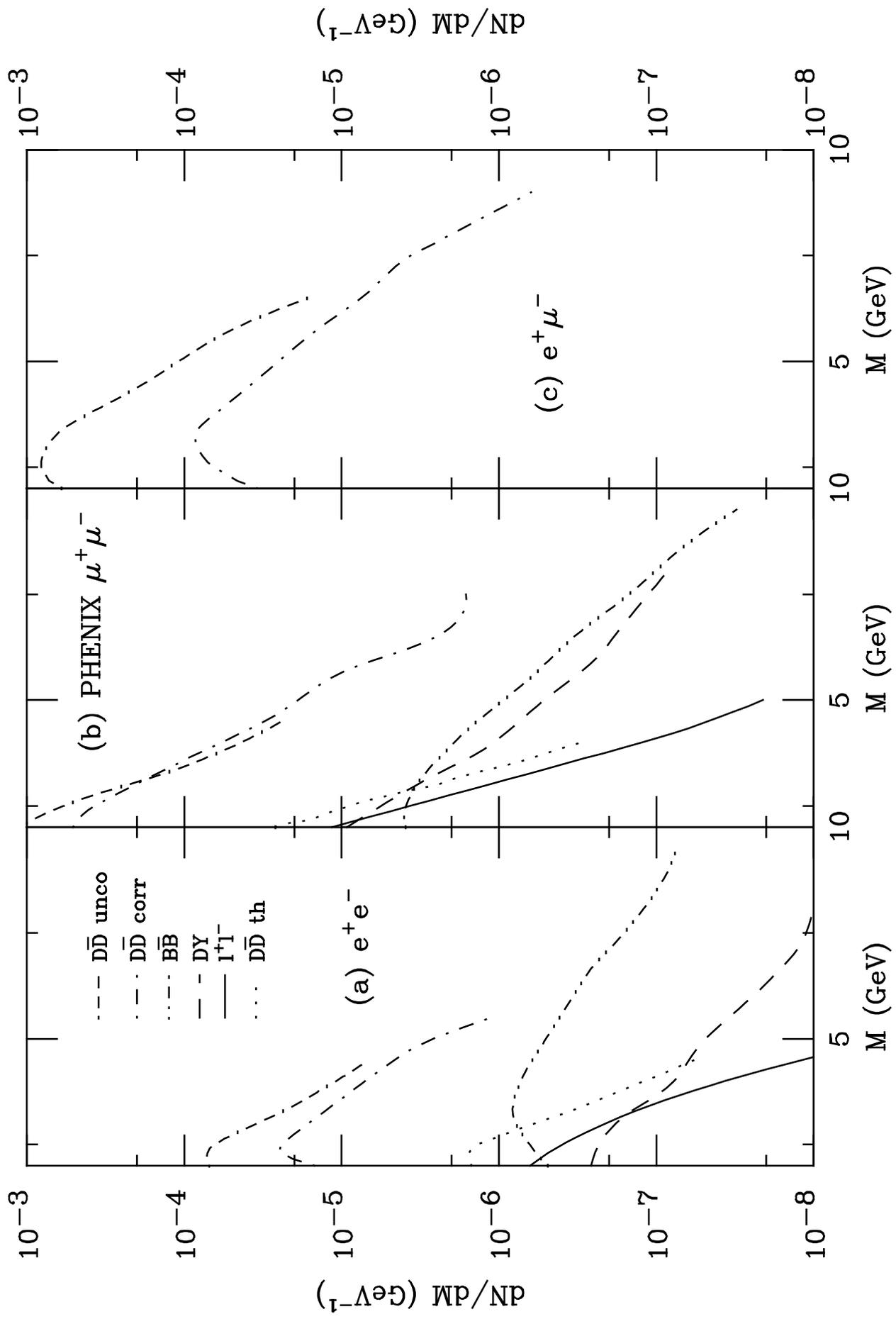
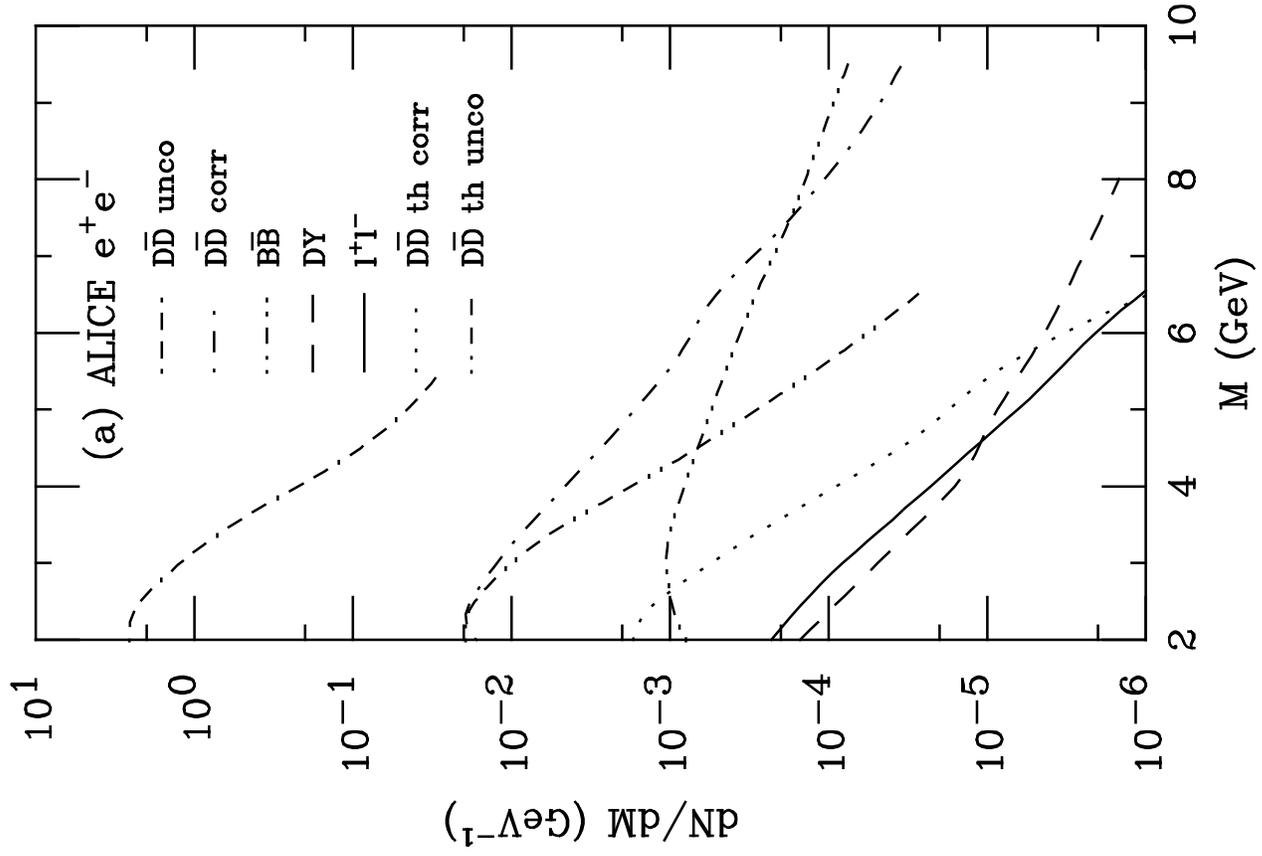
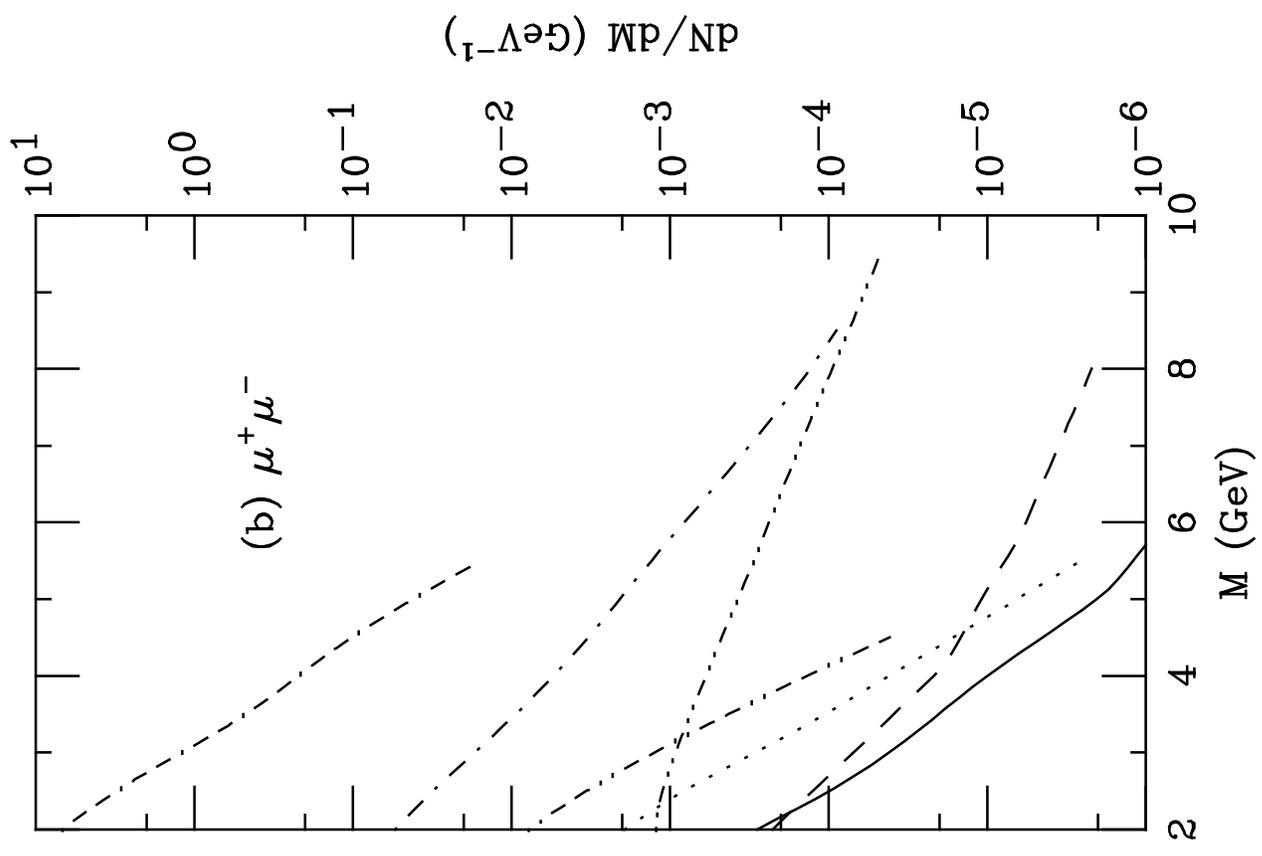
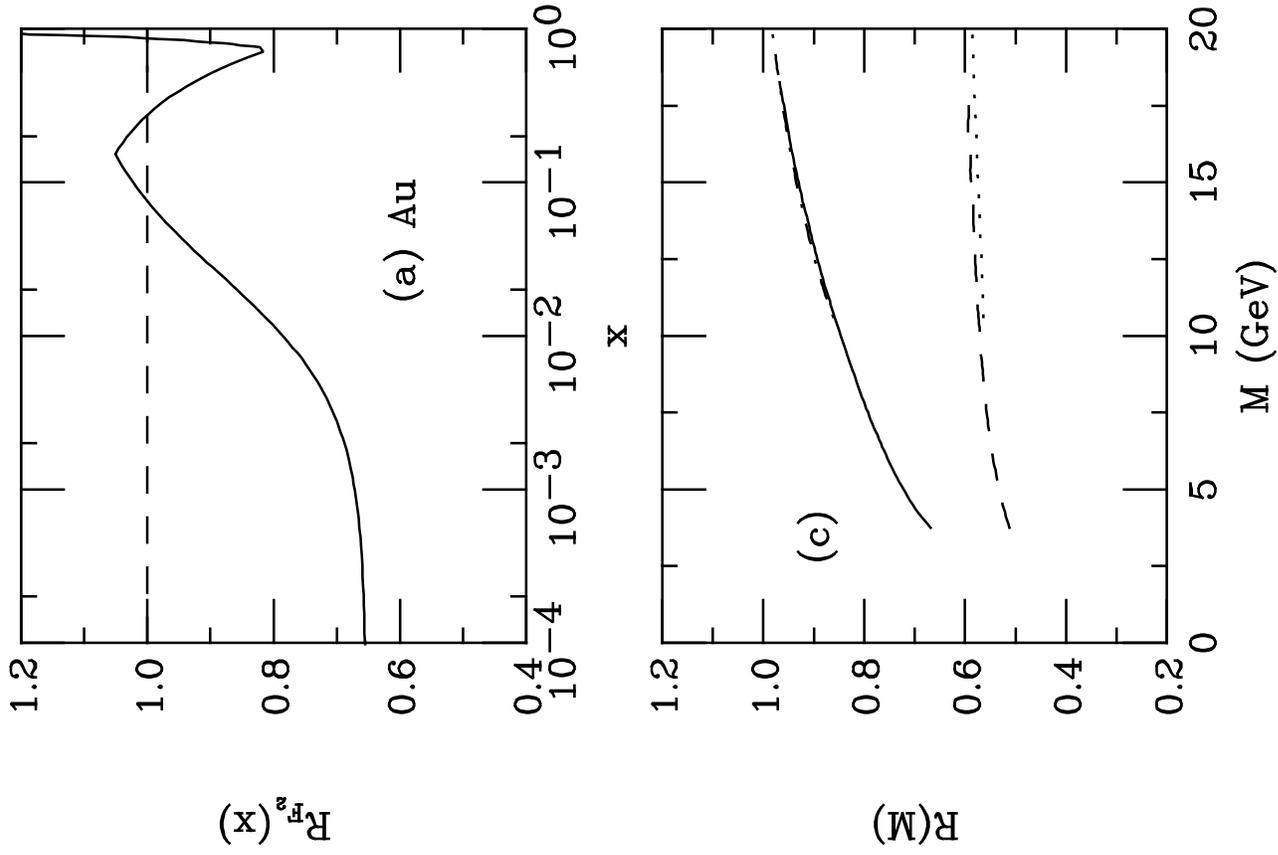
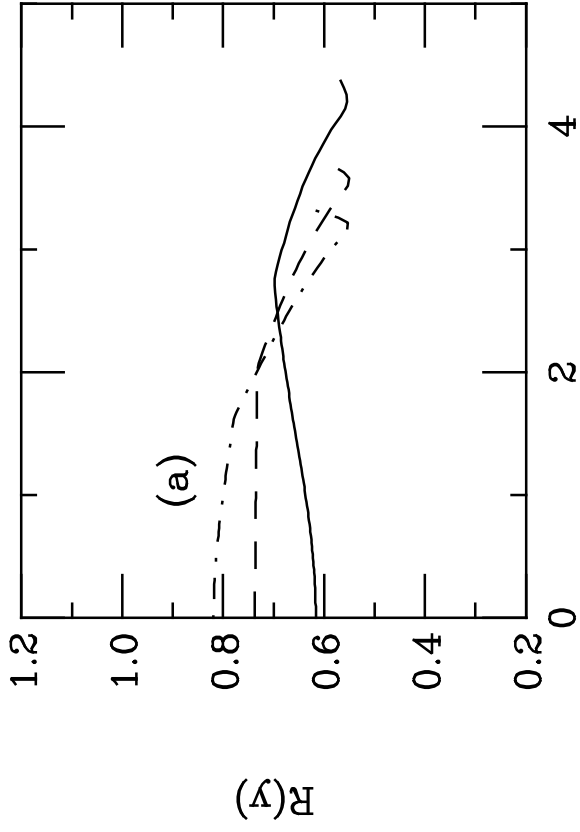
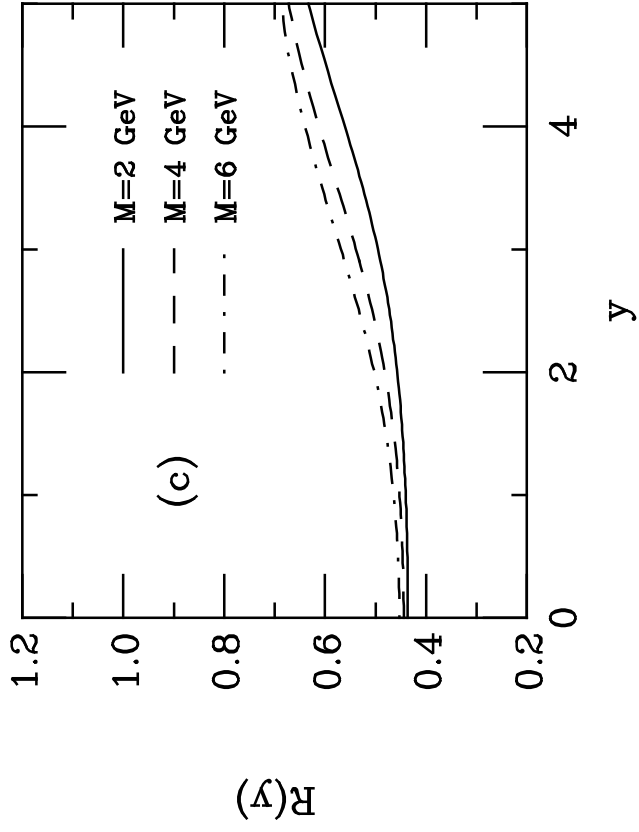
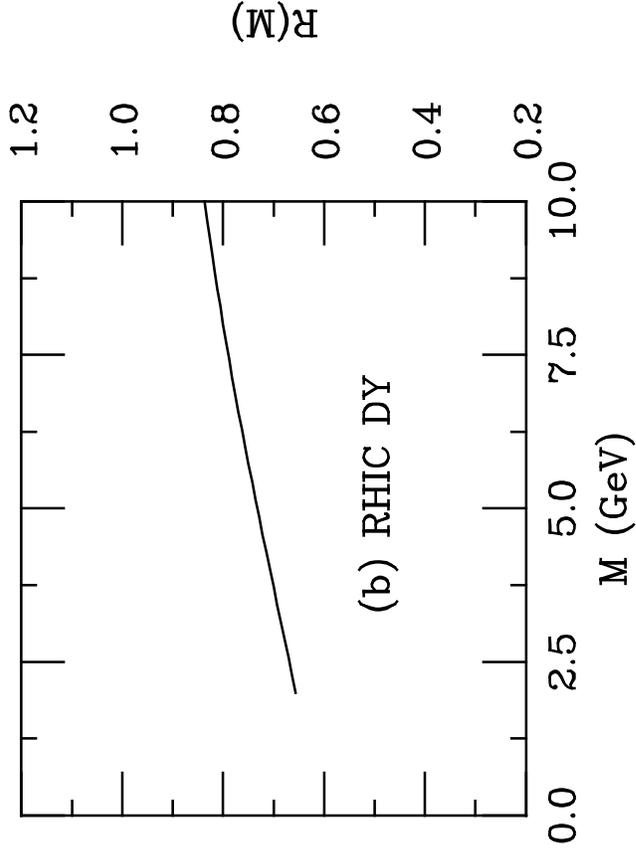
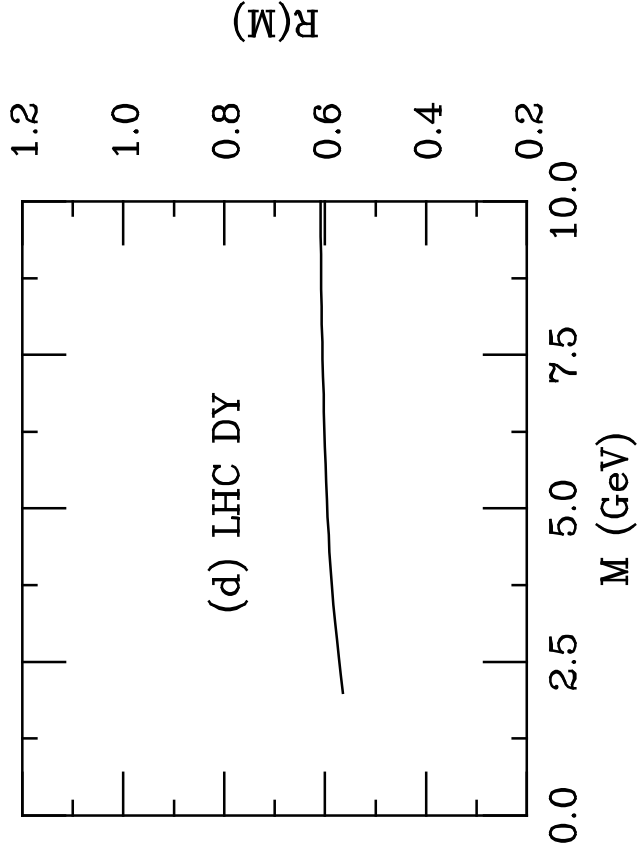
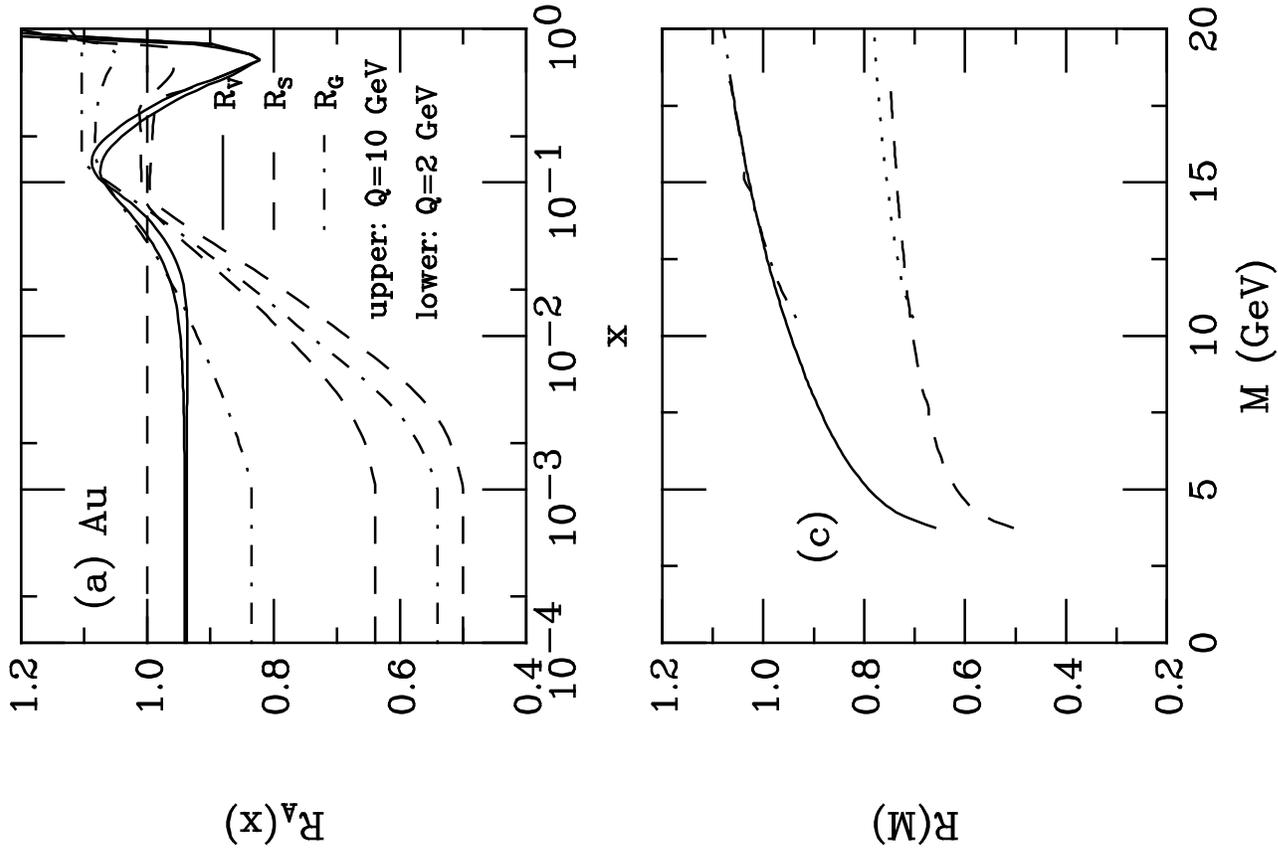
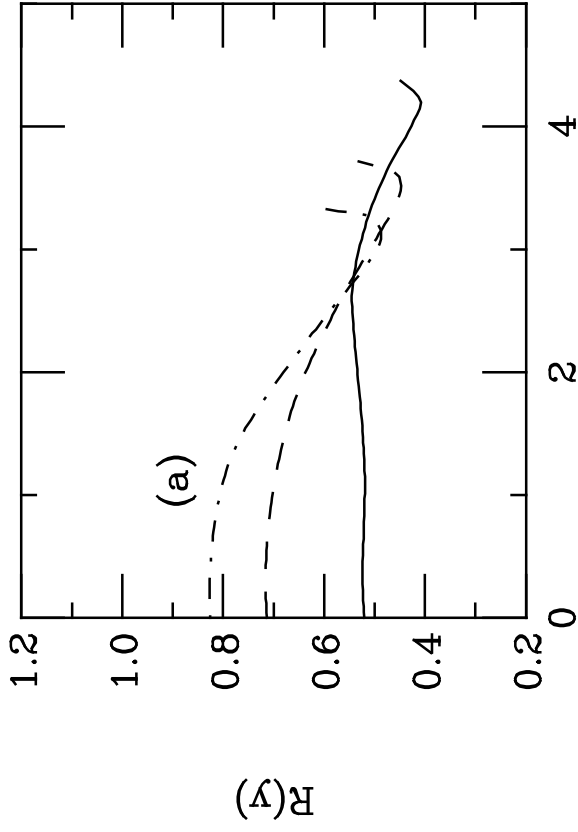
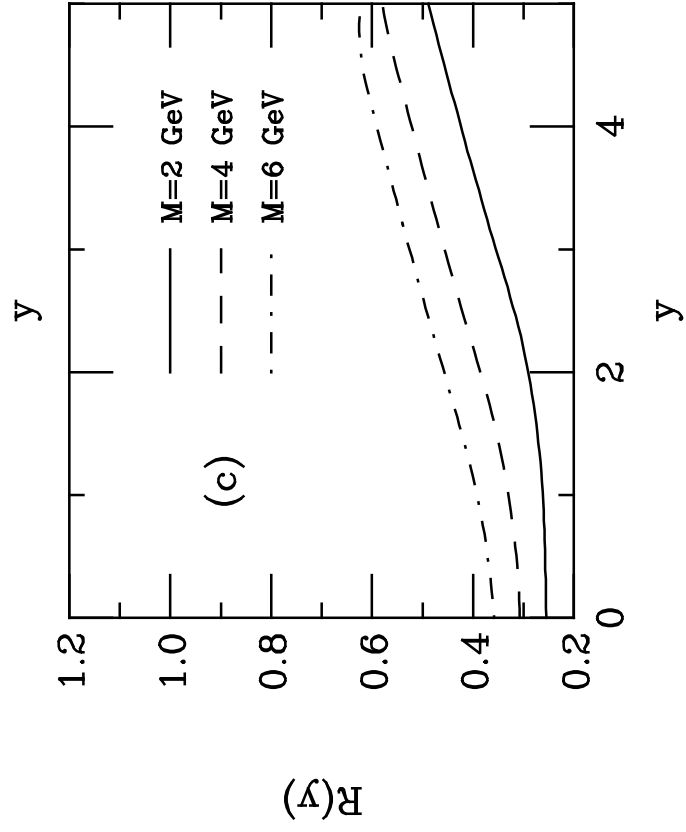
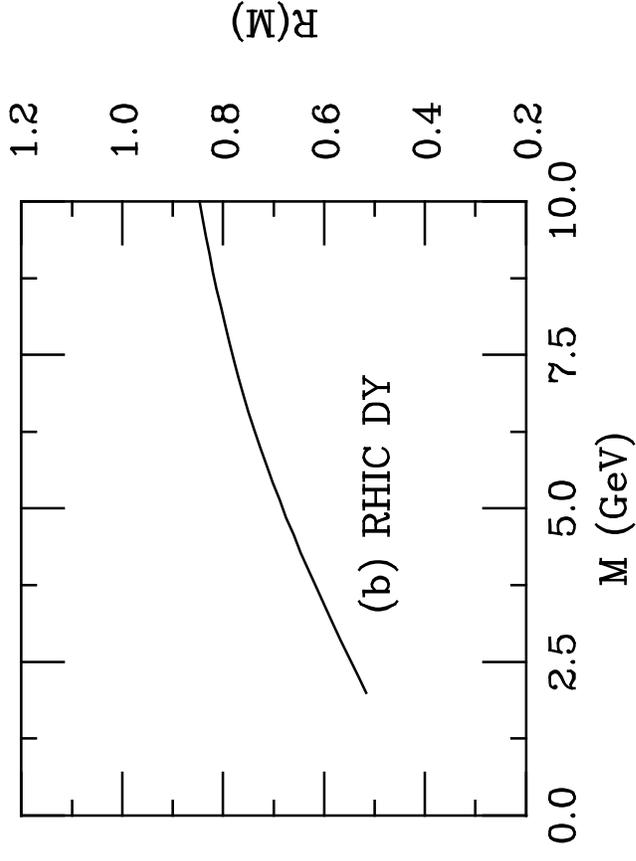
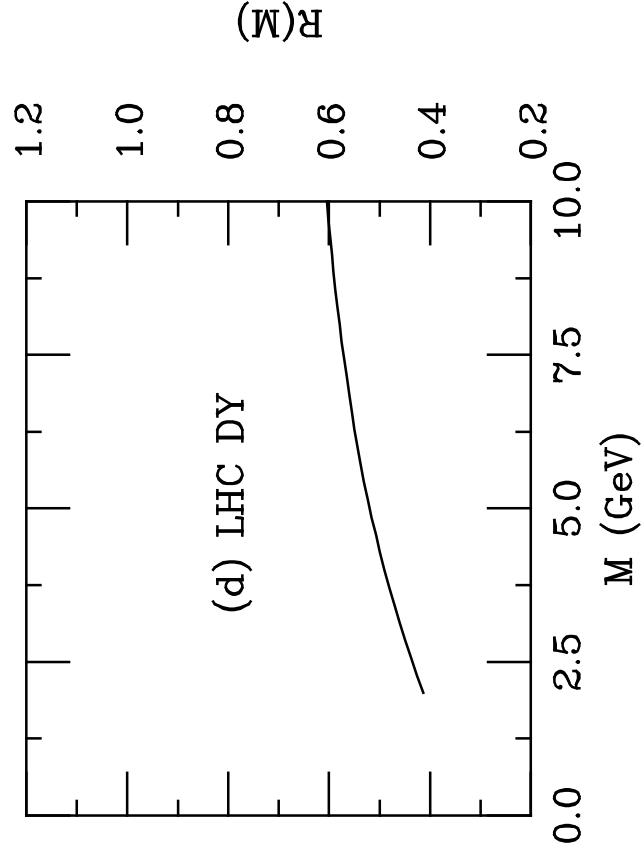
\end{document}